\documentclass[journal]{IEEEtran}
\usepackage{amsmath,bm}
\usepackage{array}
\usepackage{amssymb,amsfonts}
\usepackage{booktabs}
\usepackage[all,arc]{xy}
\usepackage{enumerate}
\usepackage{mathrsfs}
\usepackage[pass]{geometry}
\usepackage{titletoc}
\usepackage[noend]{algpseudocode}
\usepackage{graphicx}
\usepackage[caption=false,font=footnotesize]{subfig}
\usepackage{fixltx2e}
\usepackage{balance}
\usepackage{lipsum}
\usepackage{dblfloatfix} 
\usepackage{cite}
\usepackage{enumitem}
\usepackage[letterspace=150]{microtype}
\usepackage{kantlipsum}
\usepackage{siunitx}
\usepackage{booktabs}
\usepackage{amsmath}
\usepackage{array}
\usepackage{url}

\begin{document}
%
\title{Rotor Thermal Monitoring Scheme for Direct-Torque-Controlled Interior Permanent Magnet Synchronous Machines via High-Frequency Rotating Flux or Torque Injection}
%
%
%

\author{Shen~Zhang,~\IEEEmembership{Member,~IEEE,}
	Sufei~Li,~\IEEEmembership{Member,~IEEE,}
	Lijun~He,~\IEEEmembership{Senior Member,~IEEE,}\\
	Jose~A.~Restrepo,~\IEEEmembership{Senior Member,~IEEE,}
        and~Thomas~G.~Habetler,~\IEEEmembership{Fellow,~IEEE}
\thanks{S. Zhang, S. Li, and T. G. Habetler are with the Department of Electrical and Computer Engineering, Georgia Institute of Technology, Atlanta, GA 30332, USA (e-mail: shenzhang@gatech.edu, sli314@gatech.edu, thabetler@ece.gatech.edu).}
\thanks{L. He is with Electrical Systems, GE Research, Niskayuna, NY 12309, USA (e-mail: lijun.he@ge.com).}
\thanks{J. A. Restrepo is with the Universidad T\'ecnica del Norte, Ibarra 100105, Ecuador, and also with the Departamento de Electr\'{o}nica y Circuitos, Universidad Sim\'{o}n Bol\'{i}var, Caracas 1080A, Venezuela (e-mail: restrepo@ieee.org).}
}

\maketitle

\begin{abstract}
Interior permanent magnet synchronous machine drives are widely employed in electric traction systems and various industrial processes. However, prolonged exposure to high temperatures while operating can demagnetize the permanent magnets to the point of irreversible demagnetization. In addition, direct measurements with infrared sensors or contact-type sensors with wireless communication can be expensive and intrusive to the motor drive systems. This paper thus proposes a nonintrusive thermal monitoring scheme for the permanent magnets inside the direct-torque-controlled interior permanent magnet synchronous machines. By applying an external high-frequency rotating flux or torque signal to the hysteresis torque controller in the motor drive, the high-frequency currents can be injected into the stator windings. The permanent magnet temperature can thus be monitored based on the induced high-frequency resistance. The nonintrusive nature of the method is indicated by the elimination of the extra sensors and no hardware change to the existing system. Finally, the effectiveness of the proposed method is validated with experimental results.
\end{abstract}

\begin{IEEEkeywords}
Direct torque control (DTC), interior permanent magnet synchronous machines (IPMSMs), temperature estimation, rotating flux injection, torque injection.
\end{IEEEkeywords}

\IEEEpeerreviewmaketitle

\section{Introduction}
\IEEEPARstart {P}{ermanent} magnet (PM) magnetization state estimation in interior permanent magnet synchronous machines (IPMSMs) can be important for precise torque control and health prognostics \cite{PM_Demag0, PM_Temp1}. However, prolonged exposure to high temperatures while operating, especially under various types of drive-inverter initiated \cite{mcfarland2013investigation, PM_Jahns2, PM_Jahns3, PM_Jahns4} or machine winding short circuit faults \cite{PM_Habetler_thesis, PM_Habetler1, PM_Hur1}, can demagnetize the PM to the point of irreversible demagnetization, which results in degradation of the torque production and efficiency, and such damage is time consuming and costly to repair \cite{PM_Hur2, Shen_C11}. 

The impacts of PM demagnetization on various PM machine signals have been widely applied as explicit fault signatures \cite{PM_Demag_reivew1, PM_Demag_reivew2, PM_Demag_reivew3, PM_review, PM_Temp2}, as the demagnetization fault can be diagnosed by detecting noticeable signal variations and harmonic spectrum in the back-EMF \cite{PM_Briz3}, flux density (flux observers or hall effect sensors) \cite{PM_Bocker1, PM_Briz4, PM_Hall}, torque ripple \cite{PM_Zhu1}, speed harmonics \cite{PM_Speed_Har}, electrical fundamental wave models \cite{PM_Bocker2}, vibrations and acoustic noise. On the other hand, the PM magnetization state can be also revealed by monitoring its temperature, since the remnant flux and coercivity of the rare-earth PM materials will gradually deteriorate in response to the excessive heat. The PM temperature can be either measured or estimated, however, measuring its temperature once the machine is assembled is nontrivial, and existing solutions, including the infrared thermography techniques \cite{PM_infrared} with infrared light (IR) sensor/camera or the contact-type sensors through wireless transmissions, are expensive and have restricted operating environments. 

Instead of PM temperature measurement, the PM temperature can be also estimated with thermal models \cite{PM_Bocker3, PM_Inverse}, the high-frequency (HF) inductance \cite{PM_Briz_L, PM_Sul_L}, or the HF PM secondary resistance that changes with temperature, which is a byproduct of the induced magnet eddy-current loss when an alternating high-frequency magnetic field is applied to the PM \cite{PM_Zhu2, PM_2RLoss, PM_Briz1}. Due to the relatively high electrical conductivity of rare-earth magnets, the resultant eddy-current loss can be significant in the magnets. To generate the HF magnetic field required to induce the PM resistance, the high-frequency signal injection based methods have been widely employed \cite{PM_2RLoss, PM_Briz1}. In particular, extensive research has been performed on field-oriented-controlled (FOC) PM machine rotor temperature estimation with direct HF current or voltage injection \cite{PM_Briz1}. For DTC-controlled machines, however, alternative injection methods need to be developed since the current/voltage signals cannot be directly controlled in closed-loop forms \cite{IM_Lijun, Shen_C6, Shen_C7}. 

This paper thus proposes a high-frequency rotating flux and a high-frequency torque injection-based rotor thermal monitoring scheme. It first validates the principle of superimposing the appropriate torque signal for determining the HF resistances with the extracted HF voltage and current signals in the stationary reference frame. In addition, the principle of PM resistance extraction and real-time signal processing techniques are also briefly discussed. Finally, the accuracy of the proposed thermal monitoring scheme is verified with experimental results at a constant-load condition.

\section{Principle of Estimating Permanent Magnet Temperature with High-Frequency Resistance}
\begin{figure}[!t]
\centering
\includegraphics[width=2.4in]{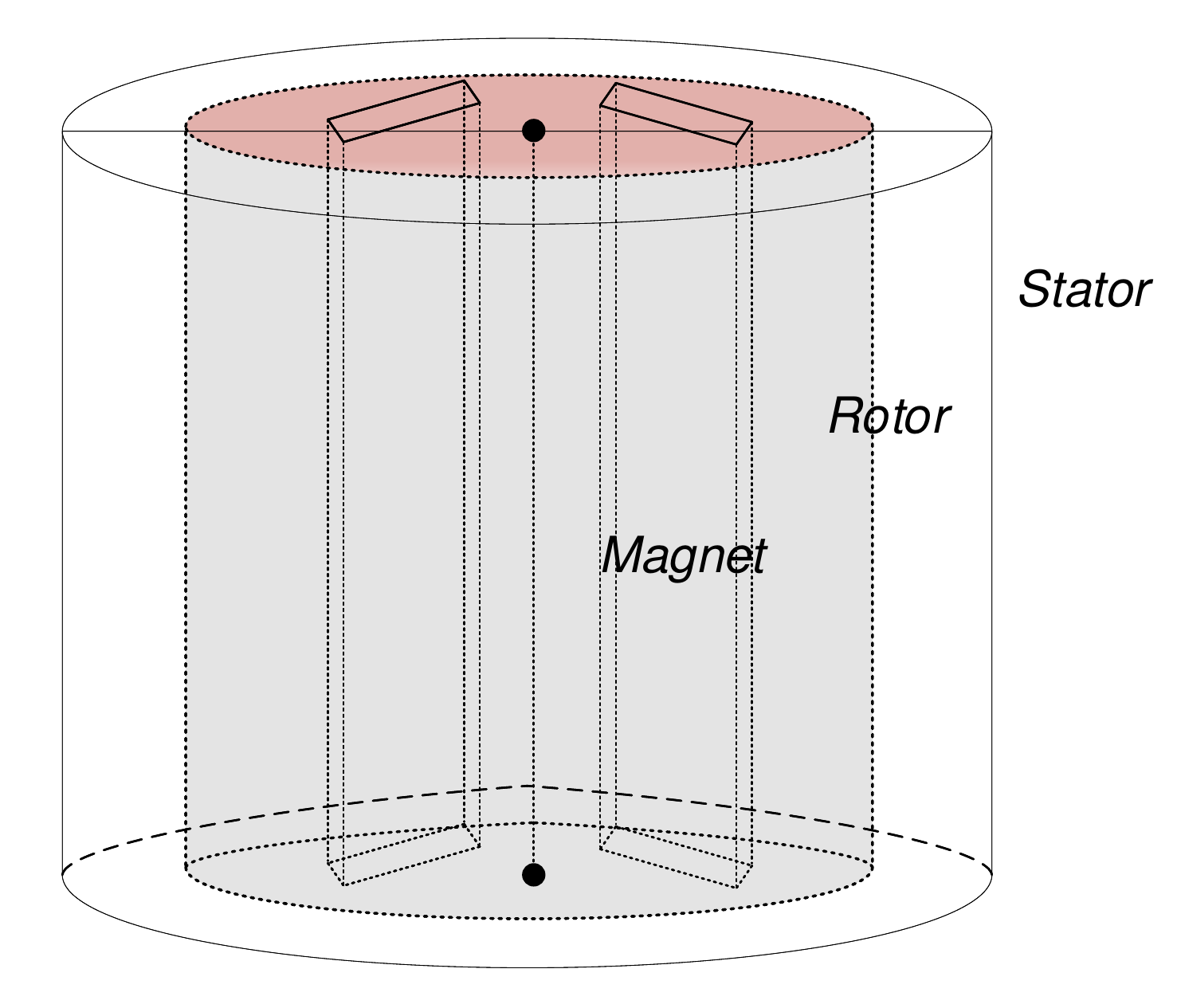}
\caption{Simple illustration of an IPMSM with ``V-shaped" magnets.}
\label{fig:IPM_V}
\end{figure}
\begin{figure}[!t]
\centering
\includegraphics[width=3.2in]{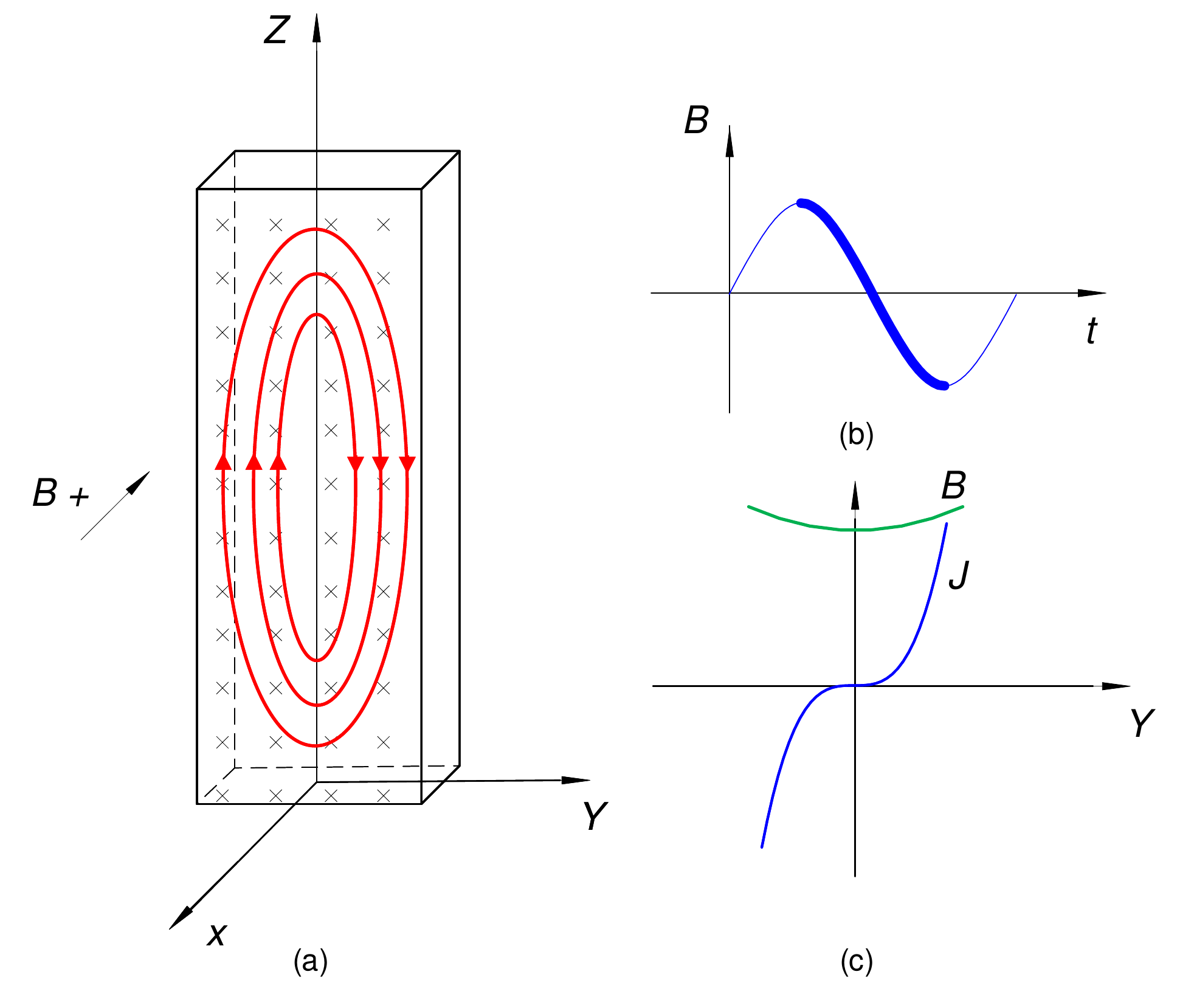}
\caption{(a) The distribution of eddy current on a permanent magnet; (b) stator flux density waveform (possible position associated with the eddy current direction) and (c) stator flux density and current density profile of a permanent magnet due to a non-predominant skin effect.}
\label{fig:IPM_eddy}
\end{figure}
Fig. \ref{fig:IPM_V} demonstrates a sketch diagram of an IPMSM, and only one pair of magnets in a single pole is plotted for simplicity. With high frequency signal injection, the induced eddy current loss per unit volume can be expressed as 
\begin{equation} 
P_{e d d y}=\frac{1}{12} \sigma \omega^{2} d^{2} B_{\alpha}^{2}
\end{equation}
where $\sigma$ is the conductivity, $d$ is the thickness of a lamination, $\omega$ is the sinusoidal frequency, and $B_{\alpha}$ is the average flux density. With a higher frequency, the eddy current loss $P_{eddy}$ increases in a quadratic manner. In addition, the skin depth can be written as
\begin{equation} 
\Delta=\sqrt{\frac{2\rho}{\omega \mu}}
\end{equation}

For the Prius IPMSM, the width of each magnet is 51.2 mm, the resistivity $\rho$ is  $\SI{1.4e-6}{\ohm}\mathrm{m}$, the permeability is $\mu= 1.05\mu_0$. Based on the above equation,  the skin depth of the $5^{th}$ order, $7^{th}$ order and $9^{th}$ order harmonics are 33.6 mm, 28.4 mm and 25.0 mm, respectively, indicating that the skin effect would be intensified by injecting higher order harmonic signals, and thus the equivalent magnet resistance reflected to the stator side would be more prominent. Fig. \ref{fig:IPM_eddy} illustrates a possible direction of the eddy current with its immediate stator flux density profile.

As discussed above, the injection of a periodic high frequency signal is a viable option for high frequency resistance estimation of a magnet. Choosing the magnitude of the high frequency signal involves a trade-off between the signal-to-noise ratio and the induced magnet losses, as larger magnitudes are advantageous for the practical implementation of the method, because of its large signal-to-noise ratio, however, it will also result in larger losses due to the eddy current and it will also distort the normal operation of the IPMSM. Therefore, the magnitude generally ranges from 2\% to 5\% of the rated value. More importantly, choosing the frequency of the high frequency signal involves a trade-off between the induced power loss and the skin effect consideration. 
\begin{figure}[!t]
\centering
\includegraphics[width=3.0in]{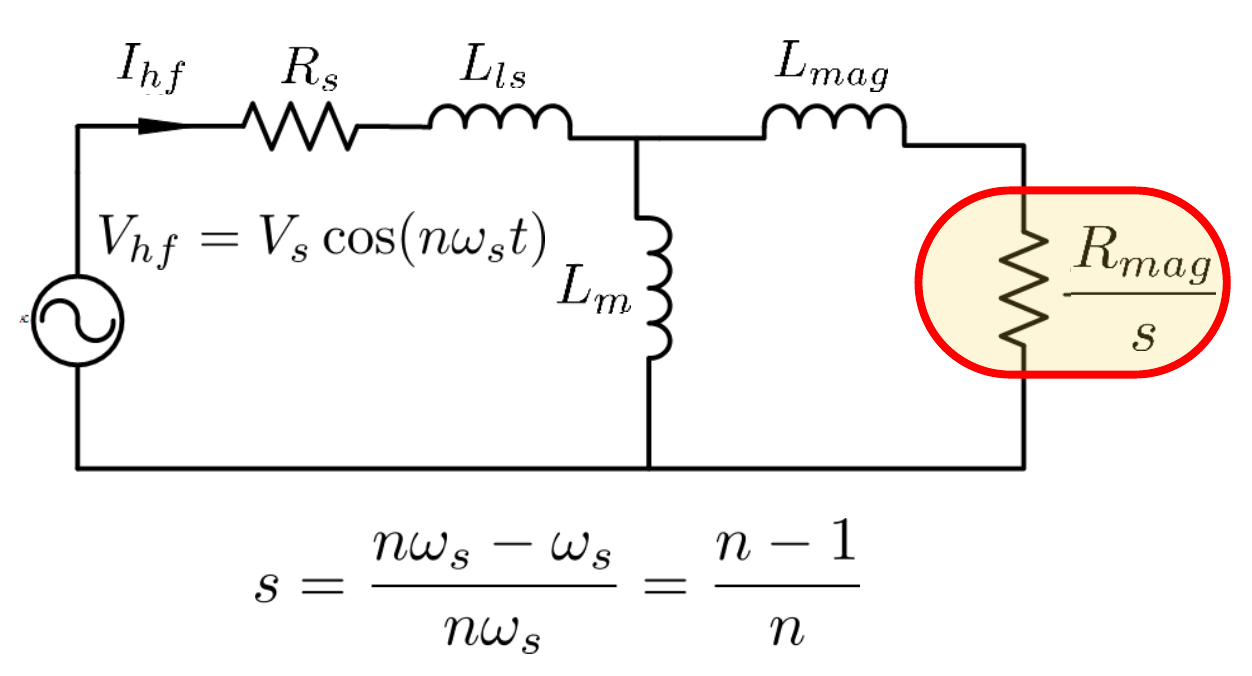}
\caption{High frequency equivalent circuit of an IPMSM at steady state with an ``equivalent slip".}
\label{fig:IPM_circuit}
\end{figure}

The magnet's resistance can be also represented in the high-frequency IPMSM equivalent circuit, which is very similar to that of an induction machine, as demonstrated in Fig. \ref{fig:IPM_circuit}, where $R_s$ and $L_{ls}$ are the stator resistance and leakage inductance, $L_m$ is the mutual inductance, $L_{lmag}$ and $R_{mag}$ are the eddy-current induced permanent magnet resistance and inductance. The slip for the high-frequency IPMSM circuit at a steady state is also defined as $n-1/n$, where $n$ is the harmonic order of the injected high-frequency signal.

It can be easily observed that when no high-frequency signal injection exists, $n = 1$, and then $s = 0$. The effective resistance of the magnet seen from Fig. \ref{fig:IPM_circuit} would become $\infty$, which means the magnet branch of Fig. \ref{fig:IPM_circuit} does not exist. This is in accordance with the fact that there is no rotor resistance or rotor loss when an IPMSM is operating at a synchronous speed. When $n$ becomes larger, the equivalent slip becomes close to 1, and then the high-frequency resistance can be obtained by extracting the high-frequency line-to-ground voltage $V_{hf}$ and current $I_{hf}$, and perform the following calculation 
\begin{equation} 
R_{hf}= \frac{\vert V_{hf}\vert}{\vert I_{hf}\vert}\cos\angle(\theta_{i}-\theta_{v}) 
\end{equation}
where $\vert V_{hf}\vert$ and $\vert I_{hf}\vert$ are the magnitude of the high-frequency component of the injected harmonic order, $\theta_v$ and $\theta_i$ are their respective phase angles. Both the current sensors and the DC bus voltage sensors are commonly available in typical motor drives. Therefore, the proposed thermal monitoring method requires no additional hardware change.

Since the calculated high-frequency resistance is a combination of the stator winding resistance $R_s$ and the PM resistance $R_{mag}$ due to the eddy current loss, the stator resistance $R_{s0}$ at room temperature $T_0$ and the real-time stator temperature $T_s$ are both required in order to decouple the PM resistance from the stator resistance. The rotor temperature can then be estimated as
\begin{equation}
T_{r}=T_{0}+ \frac{R_{hf}-R_{mag}-R_{s0}[1+\alpha_{Cu}(T_{s}-T_{0})]}{\alpha_{mag}\cdot R_{mag}} 
\end{equation}
where $\alpha_{Cu}$ and $\alpha_{mag}$ are the copper and the stator-reflected PM temperature coefficient of resistance, while $T_s$ and $T_r$ are the stator and rotor temperature.

\section{High-Frequency Resistance Extraction in DTC-Controlled IPMSMs}
\subsection{Generating the Appropriate High-Frequency Rotating Flux Linkage for Signal Injection}
In order to minimize the negative effects brought to the normal operation of IPM machines due to high-frequency signal injection, it is desirable top inject balanced three-phase high-frequency current offsets into the IPM motor three-phase windings, which can be expressed as
\begin{equation} 
\begin{bmatrix}\Delta i_{a}\\ \Delta i_{b}\\\Delta i_{c} \end{bmatrix}= \begin{bmatrix}M\cos(n\theta)\\ M\cos(n\theta-2\pi /3)\\ M\cos(n\theta+2\pi /3)\end{bmatrix}
\end{equation}
where $M$ is the magnitude of the injected high-frequency current, and $n$ is the harmonic order with respect to the fundamental frequency. Then the equivalent high-frequency current to be injected into the stationary reference frame is
\begin{equation} 
\begin{bmatrix}\Delta i_{ds}^s\\ \Delta i_{qs}^s \end{bmatrix}= \frac{2}{3}\begin{bmatrix}1 &-\frac{1}{2} &-\frac{1}{2}\\ 0 &\frac{\sqrt{3}}{2} &-\frac{\sqrt{3}}{2}\end{bmatrix} \begin{bmatrix}\Delta i_{a}\\ \Delta i_{b}\\\Delta i_{c} \end{bmatrix}=\begin{bmatrix}M \cos(n\theta)\\ M \sin(n\theta) \end{bmatrix}
\end{equation}
where the superscript $s$ denotes the stationary reference frame, $\Delta i_{ds}^s$ and $\Delta i_{qs}^s$ are the desired high-frequency current in the stationary reference frame.

The IPMSM equations in the rotor reference frame can be represented as
\begin{equation} 
\begin{bmatrix}v_{ds}^r\\ v_{qs}^r\end{bmatrix}=R_{s} \begin{bmatrix}i_{ds}^r\\ i_{qs}^r\end{bmatrix}+ \begin{bmatrix}p &-\omega_{r}\\ \omega_{r} &p\end{bmatrix} \begin{bmatrix}\lambda_{ds}^r\\ \lambda_{qs}^r\end{bmatrix} \end{equation}
where $v_{ds}^r$, $v_{qs}^r$, $\lambda_{ds}^r$ and $\lambda_{qs}^r$ are the stator voltages and flux linkages in the rotor reference frame, $R_s$ is the stator resistance, $\omega_{r}$ is the rotor speed and $p$ is the differential operator. The flux linkages can be further defined as
\begin{equation} 
\begin{bmatrix}\lambda_{ds}^r\\ \lambda_{qs}^r\end{bmatrix}= \begin{bmatrix}L_{d} &0\\ 0 &L_{q}\end{bmatrix} \begin{bmatrix}i_{ds}^r\\ i_{qs}^r\end{bmatrix}+\begin{bmatrix}\lambda_{pm}\\ 0\end{bmatrix} 
\end{equation}
in which $L_d$ and $L_q$ are the inductances for $d^r$ and $q^r$-axis, and $\lambda_{pm}$ is the flux linkage of the permanent magnet. Similarly, the IPMSM equations in the stationary reference frame is
\begin{equation} 
\begin{bmatrix}v_{ds}^s\\ v_{qs}^s\end{bmatrix} =R_{s} \begin{bmatrix}i_{ds}^s\\ i_{qs}^s\end{bmatrix}+ \begin{bmatrix}p &0\\ 0 &p\end{bmatrix} \begin{bmatrix}\lambda_{ds}^s\\ \lambda_{qs}^s\end{bmatrix} \end{equation}

By defining the transformation matrix converting the stationary reference frame to the synchronous reference frame $\mathbf{T}$ as
\begin{equation} 
\mathbf{T}=\begin{bmatrix}\cos\theta &\sin\theta\\ -\sin\theta &\cos\theta\end{bmatrix}
\end{equation}
then the flux linkage in the stationary reference frame is
\begin{gather}
 \begin{aligned} 
 \begin{bmatrix}\lambda_{ds}^s\\ \lambda_{qs}^s\end{bmatrix} = & \mathbf{T}^{-1} \begin{bmatrix}\lambda_{ds}^r\\ \lambda_{qs}^r\end{bmatrix} = \mathbf{T}^{-1} \begin{bmatrix}L_{d} & 0\\ 0 & L_{q}\end{bmatrix}\mathbf{T} \begin{bmatrix}i_{ds}^s\\ i_{qs}^s\end{bmatrix}+\mathbf{T}^{-1} \begin{bmatrix}\lambda_{pm}\\ 0\end{bmatrix}\\ = & \begin{bmatrix}\Sigma L+\Delta L\cos (2\theta) &\Delta L\sin (2\theta)\\ \Delta L\sin (2\theta) &\Sigma L-\Delta L\cos(2\theta)\end{bmatrix} \begin{bmatrix}i_{ds}^s\\ i_{qs}^s\end{bmatrix}\\&+\lambda_{pm} \begin{bmatrix}\cos \theta\\ \sin\theta\end{bmatrix}
 \end{aligned}   \raisetag{3\baselineskip}
\end{gather}
in which $\Sigma L=(L_d+L_q)/2$ and $\Delta L=(L_d-L_q)/2$. Then the small signal change of flux linkage $\Delta \lambda_{ds}^s$ and $\Delta \lambda_{qs}^s$ can be expressed as
\begin{equation} 
\begin{bmatrix}\Delta\lambda_{ds}^s\\ \Delta\lambda_{qs}^s\end{bmatrix}= \begin{bmatrix}\Sigma L+\Delta L\cos (2\theta) &\Delta L\sin(2\theta)\\ \Delta L\sin (2\theta) &\Sigma L-\Delta L\cos (2\theta)\end{bmatrix} \begin{bmatrix}\Delta i_{ds}^s\\ \Delta i_{qs}^s\end{bmatrix} \end{equation}

Substituting (6) into (12) and after performing some trigonometric transformations, the small signal model for the flux linkages can be written as
  \begin{align}
  \begin{bmatrix}\Delta \lambda_{ds}^s\\ \Delta \lambda_{qs}^s\end{bmatrix}&=\begin{bmatrix}\Sigma L\cdot M\cos(n\theta)+\Delta L\cdot M\cos[(n-2)\theta]\\ \Sigma L\cdot M\sin(n\theta)-\Delta L\cdot M\sin[(n-2)\theta]\end{bmatrix} 
   \end{align}
which yields the desired high-frequency flux linkage to be injected into the direct torque control scheme. In addition, the trajectory of this high-frequency flux linkage is a set of eclipses centering around the origin, and thus this mechanism is referred to as  the ``rotating flux linkage" injection.
 
%
\begin{figure*}[!b]
\hrule
\vspace{0.1in}
\begin{align*}
\begin{bmatrix} \Delta v_{ds}^s\\ \Delta v_{qs}^s \end{bmatrix} & =\begin{bmatrix} R_{dhf} & 0\\ 0 & R_{qhf} \end{bmatrix} \begin{bmatrix} \Delta i_{ds}^s\\ \Delta i_{qs}^s \end{bmatrix}+\begin{bmatrix} p & 0\\ 0 & p \end{bmatrix} \begin{bmatrix} \Sigma L\cdot M\mathrm{cos}(n\theta)+\Delta L\cdot M\mathrm{cos}[(n-2)\theta]\\ \Sigma L\cdot M\mathrm{sin}(n\theta)-\Delta L\cdot M\mathrm{sin}[(n-2)\theta] \end{bmatrix}\\ & = \begin{bmatrix} R_{dhf} & 0\\ 0 & R_{qhf} \end{bmatrix} \begin{bmatrix} \Delta i_{ds}^s\\ \Delta i_{qs}^s \end{bmatrix}+ \underbrace{\begin{bmatrix} -n\Sigma L\cdot M\mathrm{sin}(n\theta)p\theta\\ n\Sigma L\cdot M\mathrm{cos}(n\theta)p\theta \end{bmatrix}}_ {n^{th}\ order \ harmonic\ component} \tag{16}+\underbrace{\begin{bmatrix} -(n-2)\Delta L\cdot M\mathrm{sin}[(n-2)\theta]p\theta\\ -(n-2)\Delta L\cdot M\mathrm{cos}[(n-2)\theta]p\theta \end{bmatrix}}_{(n-2)^{th}\ order\ harmonic\ component}
\end{align*}
\end{figure*}
\begin{figure*}[!b]
\begin{align*}
\Delta T_{em}&=\displaystyle{\frac{3}{2}\frac{poles}{2}} \left\{[\Sigma L\cdot M\cos(n\theta)+\Delta L\cdot M\cos((n-2)\theta)]\cdot i_{\beta s}-[\Sigma L\cdot M\sin(n\theta)+\Delta L\cdot M\sin((n-2)\theta)]\cdot i_{\alpha s}\right. \\
&\left.+\left[\Sigma L\cdot i_{\alpha s}+\Delta L\cdot\cos(2\theta)\cdot i_{\alpha s}+\Delta L\cdot \sin(2\theta)\cdot i_{\beta s}+ \lambda_{pm}\cos\theta\right]\cdot M\sin(n\theta)\right.\\ &\left.-\left[\Delta L\cdot\sin(2\theta)\cdot i_{\alpha s}+\Sigma L\cdot i_{\beta s}-\Delta L\cdot\cos(2\theta)\cdot i_{\beta s}+ \lambda_{pm}\sin\theta\right]\cdot M\cos(n\theta)\right\}\tag{21}
\end{align*}
\end{figure*}

After injecting the flux linkage offset in (13), the high-frequency voltage and current components will appear as the system response, which can be used to calculate the high-frequency rotor resistance for estimating the PM temperature. This mechanism can be also validated in the following steps. Assume a small signal of high frequency voltage is originated under the IPM machine framework depicted in Eqn. (7), which in the stationary reference frame can be expressed as
\begin{gather}
\begin{aligned} 
\begin{bmatrix} v_{ds}^s+\Delta v_{ds}^s\\ v_{qs}^s+\Delta v_{qs}^s \end{bmatrix}=&R_{s} \begin{bmatrix} i_{ds}^s+\Delta i_{ds}^s\\ i_{qs}^s+\Delta i_{qs}^s \end{bmatrix}+\begin{bmatrix} p & 0\\ 0 & p \end{bmatrix} \begin{bmatrix} \lambda_{ds}^s+\Delta\lambda_{ds}^s\\ \lambda_{qs}^s+\Delta\lambda_{qs}^s \end{bmatrix}  \\&+\begin{bmatrix} R_{rhf\_\alpha} & 0\\ 0 & R_{rhf\_\beta} \end{bmatrix} \begin{bmatrix} \Delta i_{ds}^s\\ \Delta i_{qs}^s \end{bmatrix}  
\end{aligned} \raisetag{1.25\baselineskip}
\end{gather}

Subtracting (7) from (14), then the complete small signal model after rotating flux injection is 
\begin{equation} 
\begin{bmatrix} \Delta v_{\alpha s}\\ \Delta v_{\beta s} \end{bmatrix}= \begin{bmatrix}R_\alpha & 0\\ 0 & R_\beta \end{bmatrix} \begin{bmatrix} \Delta i_{\alpha s}\\ \Delta i_{\beta s} \end{bmatrix}+ \begin{bmatrix} p&0\\ 0&p \end{bmatrix} \begin{bmatrix} \Delta \lambda_{\alpha s}\\ \Delta \lambda_{\beta s} \end{bmatrix} 
\end{equation}
where $ R_\alpha=R_s+R_{rhf\_\alpha}$ and $R_\beta=R_s+R_{rhf\_\beta}$.

Substituting (13) into (15), the model becomes a form shown in Eqn. (16), which contains the frequency component of both the $n^{th}$ and the $(n-2)^{th}$ order. With appropriate signal processing techniques, both frequency components should be accurately extracted from the real-time current and voltage measurements. However, since the amplitude of the $n^{th}$ harmonic order is larger and more distinguishable compared to the $(n-2)^{th}$ order, the $n^{th}$ order components should be utilized for estimating the high-frequency rotor resistance $R_r$ and thus the PM temperature.
%

%
\begin{figure*}[!t]
\centering
\includegraphics[width=6.0in]{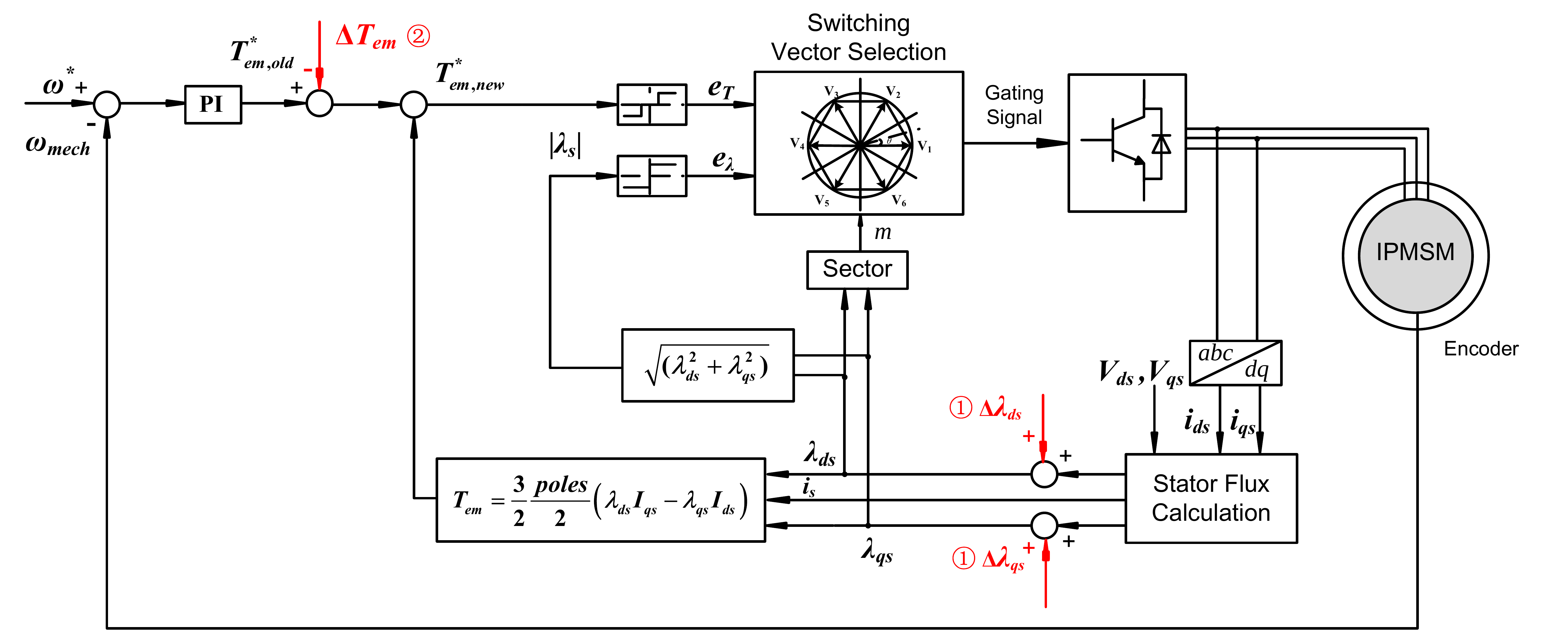}
\caption{Block diagram of the proposed high-frequency torque injection method on a conventional DTC scheme.}
\label{fig_DTC}
\end{figure*}
%
%
\begin{figure*}[!t]
\centering
\subfloat[]{\includegraphics[width=5.6in]{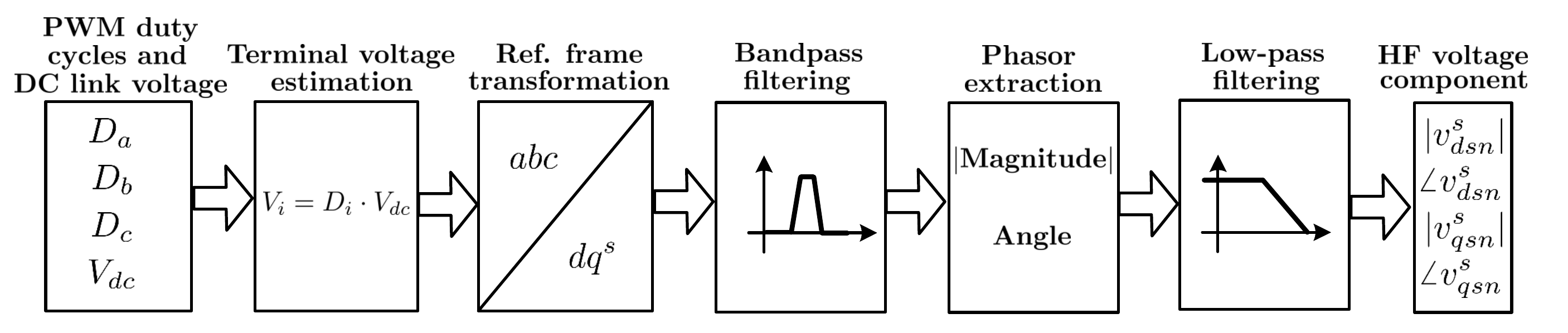}} \\
\vspace{-0.16in}
\subfloat[]{\includegraphics[width=3.6in]{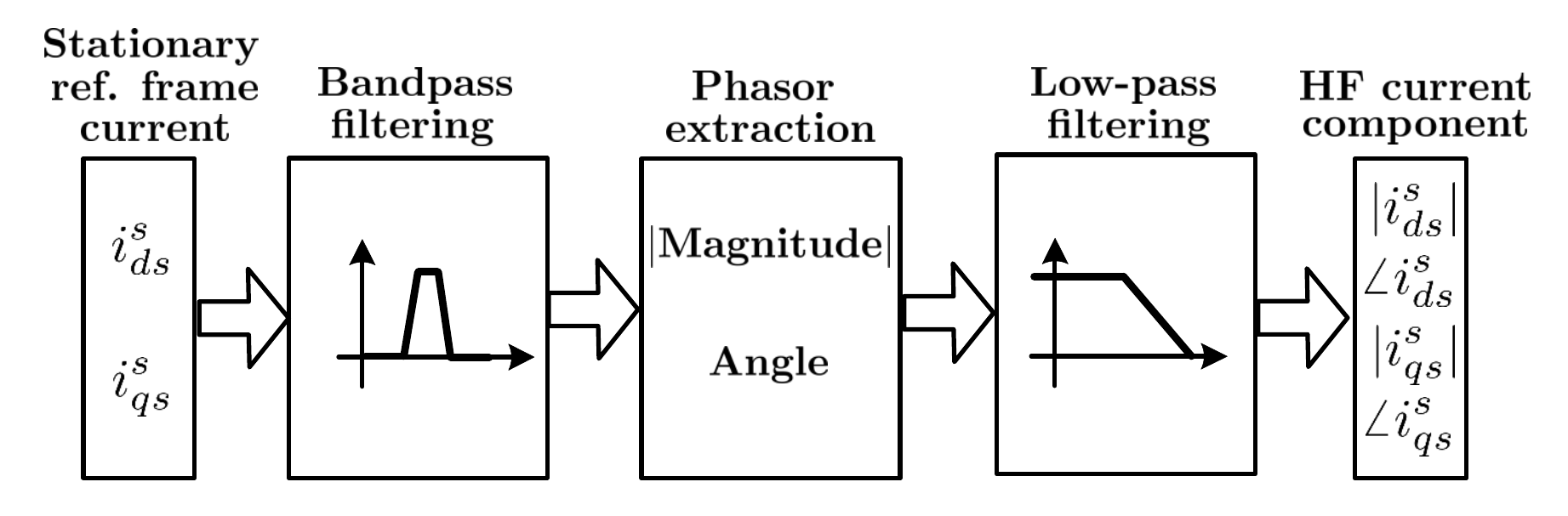}} \\
\caption{Signal processing techniques for extracting high-frequency current and voltage.} 
\label{fig_HF_extraction} 
\end{figure*} 

Since $p\theta = \omega_r$ and define $\omega_n=n\omega_r$, Eqn. (16) can be updated by only taking its $n^{th}$ order component as
 \begin{gather}
 \begin{aligned} 
 \begin{bmatrix} \Delta v_{dsn}^s\\ \Delta v_{qsn}^s \end{bmatrix}=&\begin{bmatrix} R_{d} & 0\\ 0 & R_q \end{bmatrix} \begin{bmatrix} \Delta i_{ds}^s\\ \Delta i_{qs}^s \end{bmatrix}+\begin{bmatrix} - \omega_{n} \Sigma L \cdot M \sin ( n \theta )\\  \omega_{n} \Sigma L \cdot M \cos ( n \theta ) \end{bmatrix}  \\&+\begin{bmatrix} R_{rhf\_\alpha} & 0\\ 0 & R_{rhf\_\beta} \end{bmatrix} \begin{bmatrix} \Delta i_{ds}^s\\ \Delta i_{qs}^s \end{bmatrix}  
 \end{aligned} \tag{17} \raisetag{1.25\baselineskip}
 \end{gather}
where the $n^{th}$ order high-frequency voltages are $v_{dsn}^s$ and $v_{qsn}^s$. Substituting (6) into (17) yields
 \begin{equation*}
 \begin{bmatrix} \Delta v_{dsn}^s\\ \Delta v_{qsn}^s \end{bmatrix} = \begin{bmatrix} R_{d} & 0\\ 0 & R_q \end{bmatrix} \begin{bmatrix} \Delta i_{dsn}^s\\ \Delta i_{qsn}^s \end{bmatrix}+ \omega_n \Sigma L \begin{bmatrix} -\Delta i_{qsn}^s\\  \Delta i_{dsn}^s \end{bmatrix}  \\
 \tag{18} 
 \end{equation*} 

The above equation can be further simplified into
\begin{equation*}
    \mathbf{\Delta V_n} = \mathbf{R}+ \mathbf{J}\cdot \omega_n \Sigma L \cdot \mathbf{\Delta I_n}   \\
\end{equation*}
where the matrices are defined as follows:
\begin{equation*}
    \begin{aligned}
    \mathbf{\Delta V_n} & = \begin{bmatrix} \Delta v_{dsn}^s ~~ \Delta v_{qsn}^s\end{bmatrix}^T\\
    \mathbf{R} & = \begin{bmatrix} R_d ~~ R_q \end{bmatrix}^T\\
    \mathbf{\Delta I_n} & = \begin{bmatrix} \Delta i_{dsn}^s ~~  \Delta i_{qsn}^s \end{bmatrix}^T\\
    \mathbf{J} & = \begin{bmatrix} 0 & -1 \\ 1 & 0 \end{bmatrix}.
    \end{aligned}
\end{equation*}

The above equation is the very familiar $R-L$ circuit equation. Therefore, injecting the small signal change of the high frequency flux linkage in Eqn. (8) into the DTC control scheme can produce the desired high frequency voltage and current signals for calculating the high-frequency resistance, and consequently the PM temperature.

\subsection{Generating the Appropriate High-Frequency Torque for Signal Injection}
Since DTC has direct control on both the flux linkage and the electromagnetic torque, which enables the mechanism of injecting a high-frequency torque signal besides the aforementioned flux linkage signals. Assume a small change is superimposed on the torque same as the torque injection $\Delta T_{em}$ at a high frequency (e.g., $5^{th}$) equation as expressed
\begin{align*} 
T_{em}+ \Delta T_{em}&=\frac{3}{2}\frac{poles}{2}[(\lambda_{\alpha s}+\Delta\lambda_{\alpha s})(i_{\beta s}+\Delta i_{\beta s})\\&-(\lambda_{\beta s}+\Delta\lambda_{\beta s})(i_{\alpha s}+\Delta i_{\alpha s})]\tag{19}  
\end{align*}

Subtracting it from the original $T_{em}$ signal while neglecting the second-order small signal approximations. Then the small change of torque can be written as
\begin{align*} \Delta T_{em}=\frac{3}{2}\frac{poles}{2}(&\Delta\lambda_{\alpha s}i_{\beta s}+\lambda_{\alpha s}\Delta i_{\beta s}\\&-\Delta\lambda_{\beta s}i_{\alpha s}-\lambda_{\beta s}\Delta i_{\alpha s})\tag{20}  \end{align*}

Finding a simplified equation for the torque offset to be superimposed is a nontrivial task, since substituting Eqn. (13) into the above equation yields Eqn. (21) shown below.

After performing some rigorous trigonometric derivation processes, the above equation can be greatly simplified into
\begin{align*} \Delta T_{em}&=\frac{3}{2}\frac{poles}{2}P\{2\Delta Li_{\beta s}\cos[(n-2)\theta] \\&+2\Delta Li_{\alpha s}\sin[(n-2)\theta]+\lambda_{pm}\sin [(n-1)\theta]\}  \tag{22}  \end{align*}

\subsection{Illustration of the Proposed High-Frequency Torque Injection Technique}
There are several important implications of the proposed torque injection signal in Eqn. (22) that can be summarized as follows, for the first term, since $i_{\beta s}$ is a sine-wave current signal at the fundamental frequency (1st order), while getting multiplied by another $(n-2)^{th}$ order signal will yield a combination of a $(n-1)^{th}$ and a $(n-3)^{th}$ order signal. Since the second term in (12) is also of $(n-1)^{th}$ order, the expression in (12) indicate that a combination of $(n-1)^{th}$ and $(n-3)^{th}$ order varying torque command needs to be superimposed on the original torque reference to practically inject the high-frequency current component, while no modification for the flux control loop is required. Although the inner DTC loop for a typical IPM motor generally does not have a bandwidth that is capable of reinforcing the high-frequency torque bias to be exactly the same content, the actual system response can still be kept stable at a constant magnitude at the desired high frequency, which is sufficient to trigger the high-frequency current and voltage signals.

The block diagram of this injection method based on the conventional DTC scheme is demonstrated in Fig. \ref{fig_DTC}. Similar to the earlier discussion, although the bandwidth of the speed loop is typically too low to completely cancel out $\Delta T_{em}$, the output of the speed regulator $T_{em,old}$ will still be varying sinusoidally and partially compensating the external injection $\Delta T_{em}$, due to the small speed ripple. As a result, the magnitude of the actual torque ripple in $T_{em,new}$ is reduced, and the phase is shifted. However, this torque bias is still a combination of sine waves varying at the $(n-1)^{th}$ and $(n-3)^{th}$ order, but with a different combination of magnitude and phase angle, compared with the external injection signal in (22). In addition, this sine-wave torque injection does result in a small amount of extra torque ripple. However, considering the inertia of the rotor and load, the speed ripple is almost unnoticeable. Moreover, the injection technique is only required to be implemented for a short period of time, like 20 seconds once every 10 minutes; and the speed ripple can be further limited by choosing a small enough $\Delta i_{ds} (P)$. To summarize, the impact of this high-frequency signal injection on the normal operation of the IPM motors is almost negligible.
\begin{figure*}[!t]
\centering
\subfloat[]{\includegraphics[width=5.6in]{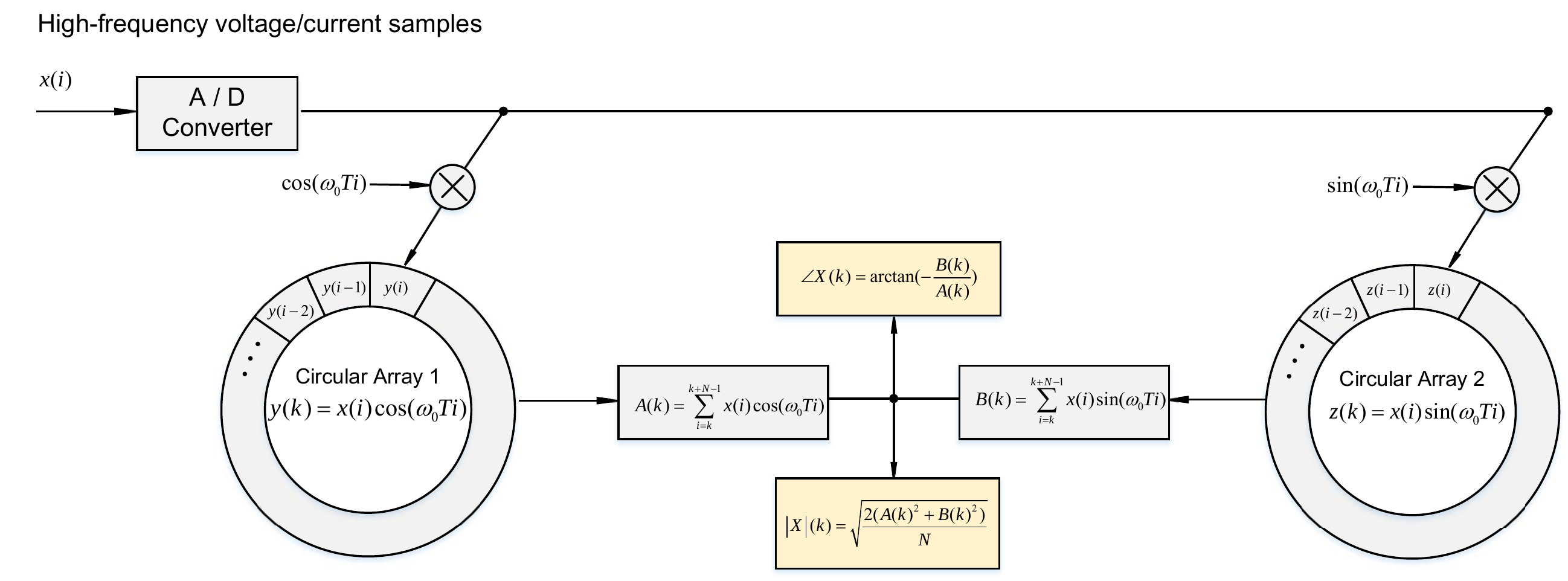}}
\caption{Signal processing techniques for extracting high-frequency current and voltage.} 
\label{fig_signal} 
\end{figure*} 
\section{Real-Time Signal Processing Techniques}
As demonstrated in Fig. \ref{fig_signal}(a) and \ref{fig_signal}(b), two second-order Butterworth bandpass filters are designed to extract the high-frequency voltage and current signals. Assume that the measured high-frequency voltage and current signals are uniformly sampled at a rate of $N$ samples per period, the phasor extraction computations can be implemented with high efficiency using a circular array based algorithm \cite{ECE6323}, then computations such as moving averages and Discrete Fourier transforms can be implemented in this way.

This phasor extraction algorithm starts from setting up two circular arrays with N entries each with all zero initial values. Then two accumulators defined as $A(k)$ and $B(k)$ are also set to zeros. As each data sample $x(i)$ arrives, the following steps are performed

\subsubsection{Compute the two temporary values updating the following sums recursively at each new available data sample}
\begin{align*} &y(k)=x(i)\cos(\omega_{0}Ti)\\ &z(k)=x(i)\sin(\omega_{0}Ti) \end{align*}
where $\omega_0$ is the base frequency, $T$ is the sampling period.
\subsubsection{Update the two accumulators as}
\begin{align*} &A(k)=A(k-1)+y(i)-y(i-N)\\ &B(k)=B(k-1)+z(i)-z(i-N) \end{align*}
where $N$ is the sampling rate and is equal to $2\pi/\omega_0T$.
\subsubsection{Overwrite the $N$-step-previous values $y(i-N)$ and $z(i-N)$ with the present values of $y(i)$ and $z(i)$ in the arrays. The above expressions have more compact forms as}
\begin{align*} &A(k)= \sum\limits_{i=k}^{k+N-1}x(t)\cos(\omega_{0}Ti)\\ &B(k)= \sum\limits_{i=k}^{k+N-1}x(i)\cos(\omega_{0}Ti) \end{align*}
\subsubsection{Compute the magnitude and angle at the present step}
\begin{gather*} \vert X(k)\vert =\sqrt{\frac{2(A(k)^{2}+B(k)^{2})}{N}}\\ \angle X(k)= \arctan (- \frac{B(k)}{A(k)}) \end{gather*}

Finally, the computed voltage/current magnitude and phase angle values will then pass through second-order low pass filters to eliminate any remaining high-frequency harmonics, and thus calculate the high-frequency resistance as in Eqn. (3). 
\section{Experimental Validation}
\subsection{Experiment Setup}
\begin{figure}[!t]
\centering
\subfloat[]{\includegraphics[width=3.0in]{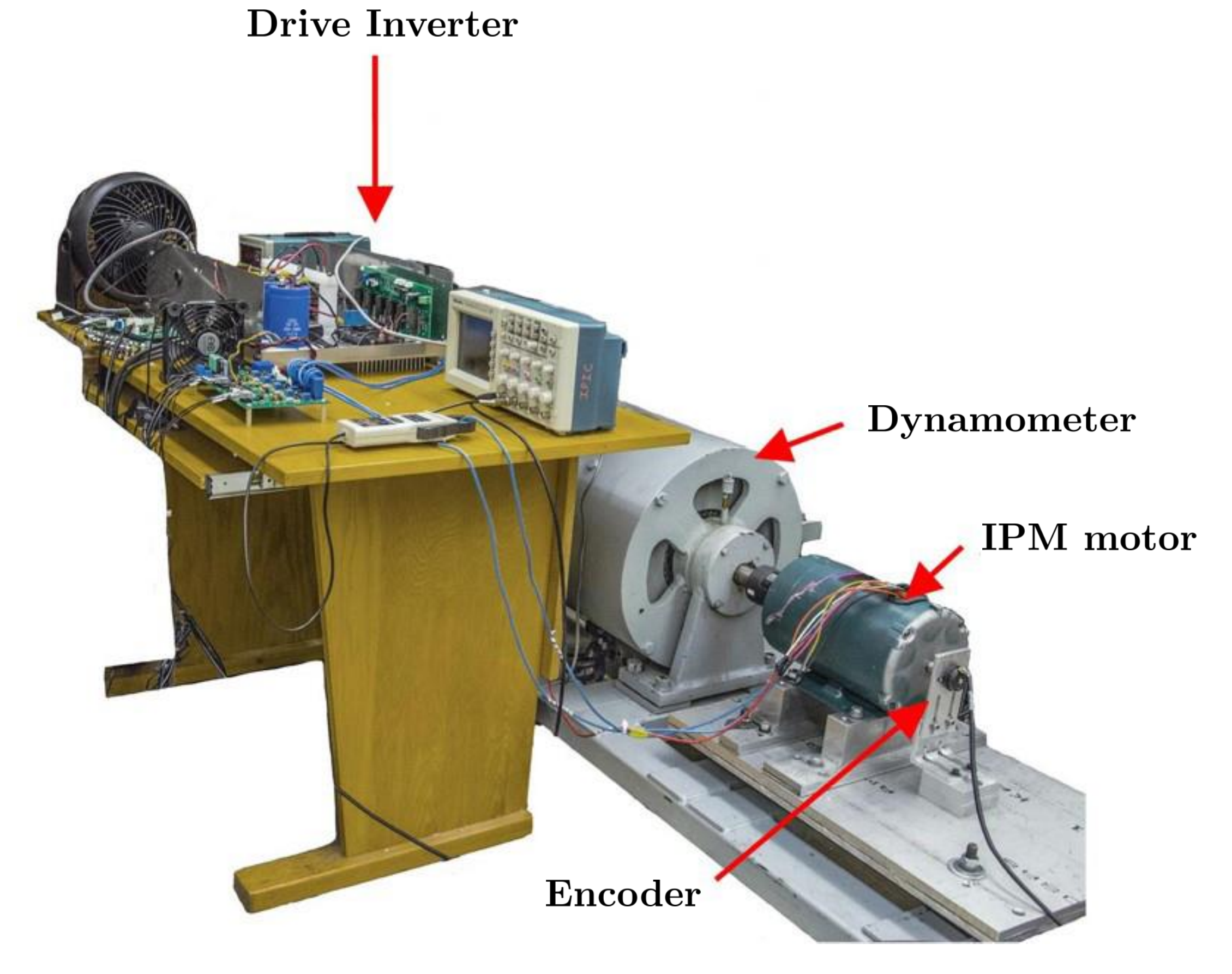}} \\
\subfloat[]{\includegraphics[width=1.8in]{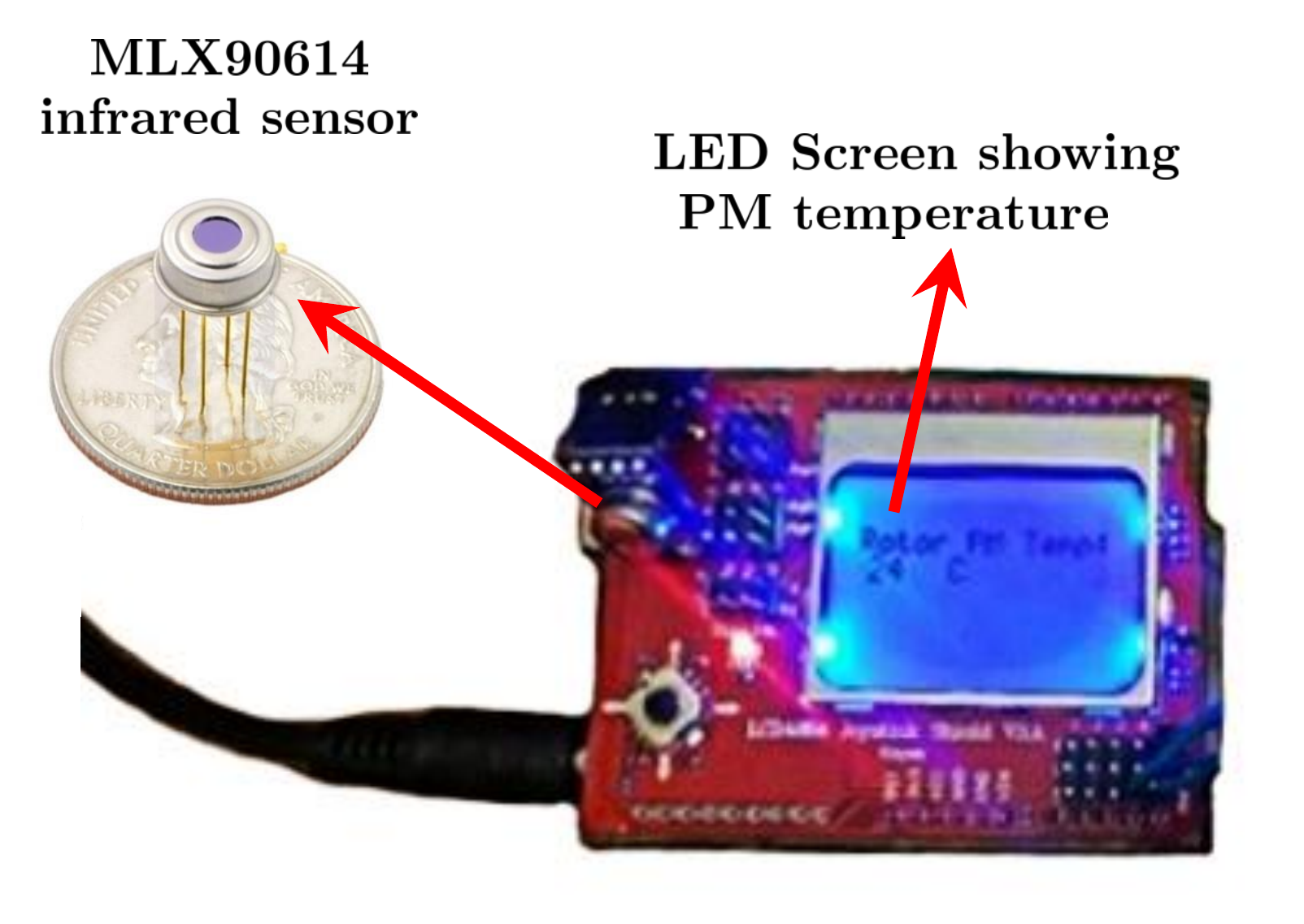}} 
\caption{(a) Experimental setup for the proposed rotor thermal monitoring method and (b) displaying the measured rotor temperature with an Arduino board.} 
\label{fig:PM_setup} 
\end{figure}

A programmable inverter drive was designed and built to validate the proposed thermal monitoring method, and the entire hardware system is illustrated in Fig. \ref{fig:PM_setup}. An Analog Device 21369 DSP is used in conjunction with a Xilinx Spartan-3 field-programmable gate array (FPGA) as the motor controller. The inverter runs at 20 kHz PWM frequency. A direct torque control scheme is implemented in the inverter drive and the proposed thermal monitoring method is integrated in the control algorithm. The parameters of the IPM motor are presented in TABLE \ref{tab_I}.

The stator temperature $T_s$ is measured with four K-type thermocouples attached to the stator winding, while the reference rotor temperature is measured with the Melexis MLX90614 infrared sensor through two measuring holes drilled on the motor end cap. The sensor is controlled with an Arduino UNO board installed with a LCD screen demonstrating the real-time value of the measured PM temperature, as shown in Fig. \ref{fig:PM_setup}(b). 

\begin{table}[]
\centering
    \label{tab_I}
    \caption{Parameters of the IPM Machine.}
\begin{tabular}{ll}
\toprule
 Parameters & Value \\ 
\midrule
    Rated power $(P_{rated})$ & $1~hp$  \\                    
    Number of poles & $4$  \\ 
    Rated voltage $(V_{rated})$ & $230~V$  \\ 
    Rated current $(I_{rated})$ & $2.86~A$  \\ 
    Stator resistance $R_s$ & $\SI{2.85}{\ohm}$  \\
    d-axis unsaturated inductance $(L_d)$ & $14.41~mH$  \\ 
    q-axis unsaturated inductance $(L_q)$ & $27.92~mH$  \\ 
    Rated speed $(\omega_{rated})$ & $1,800~rpm$ \\
    Inertia $(J)$ & $0.0050~kg\cdot m^2$ \\
\bottomrule
\end{tabular}
\end{table}
\begin{figure}[!t]
\centering
\subfloat[]{\includegraphics[width=2.25in]{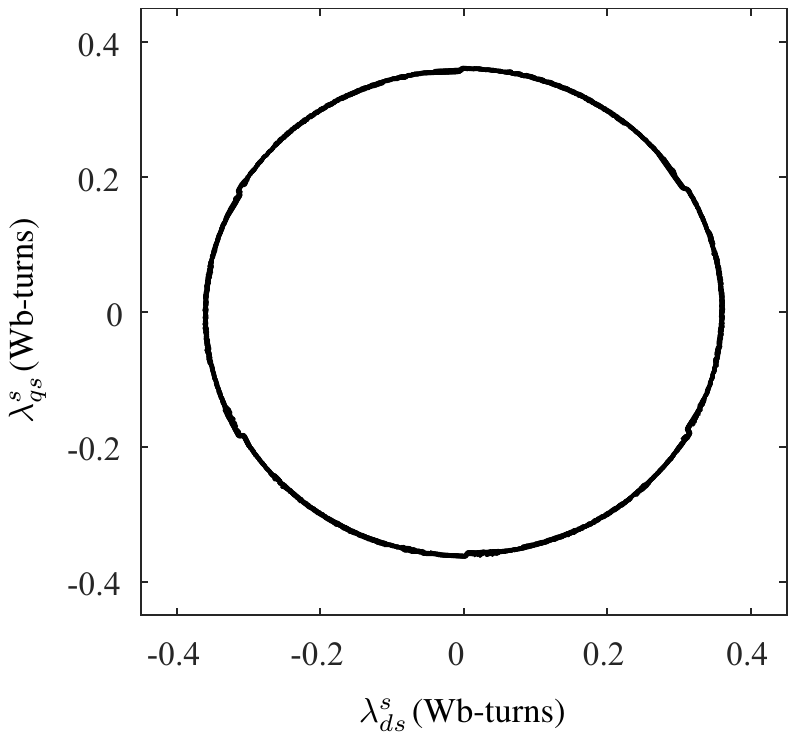}}
\hspace{0.05in}
\subfloat[]{\includegraphics[width=2.25in]{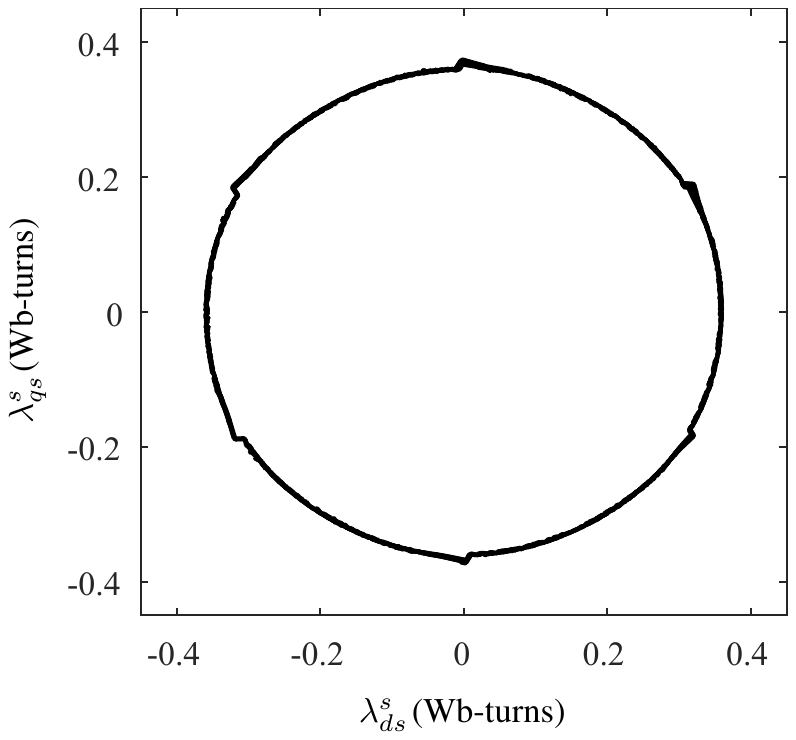}}
\caption{2-D flux linkage diagram (a) before high-frequency rotating flux injection and (b) after high-frequency rotating flux injection.} 
\label{fig:PM_flux} 
\end{figure} 

\subsection{High Frequency Voltage, Current and Flux Linkage Waveform}
The 2-D flux linkage waveform of $\lambda_{ds}^s$ against $\lambda_{qs}^s$ is demonstrated in Fig. \ref{fig:PM_flux}. Before the injection of the high-frequency signal, it can be observed in Fig. \ref{fig:PM_flux}(a) that the trajectory is almost a perfect circle with its center in the origin, which indicates that the flux linkage of the IPM machine is well-regulated and there is very little intrinsic DC offset or high-frequency harmonics. After manually injecting the rotating high-frequency flux linkage, a noticeable distortion on the circle can be observed that clearly demonstrates the high-frequency contents have been added onto the original circular trajectory.

\begin{figure*}
\centering
\subfloat{\includegraphics[width=2.2in]{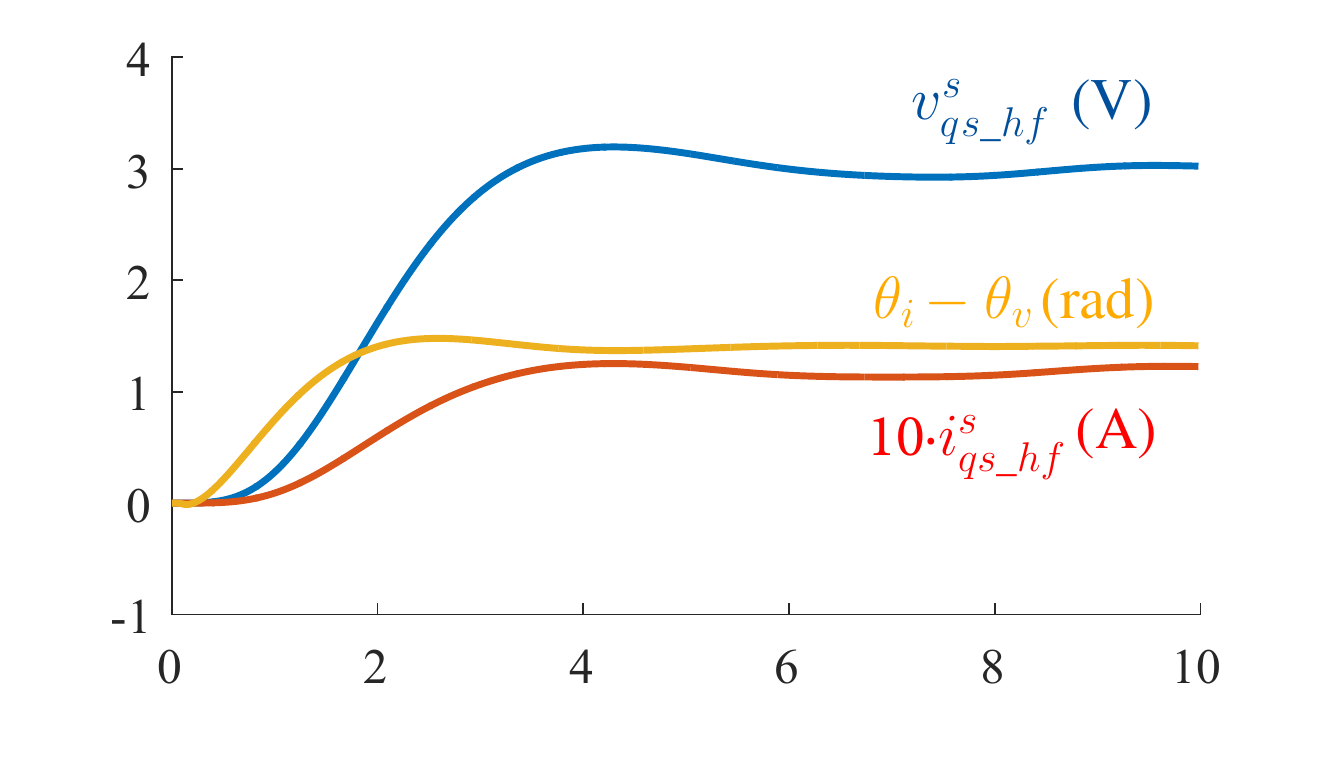}} 
\subfloat{\includegraphics[width=2.2in]{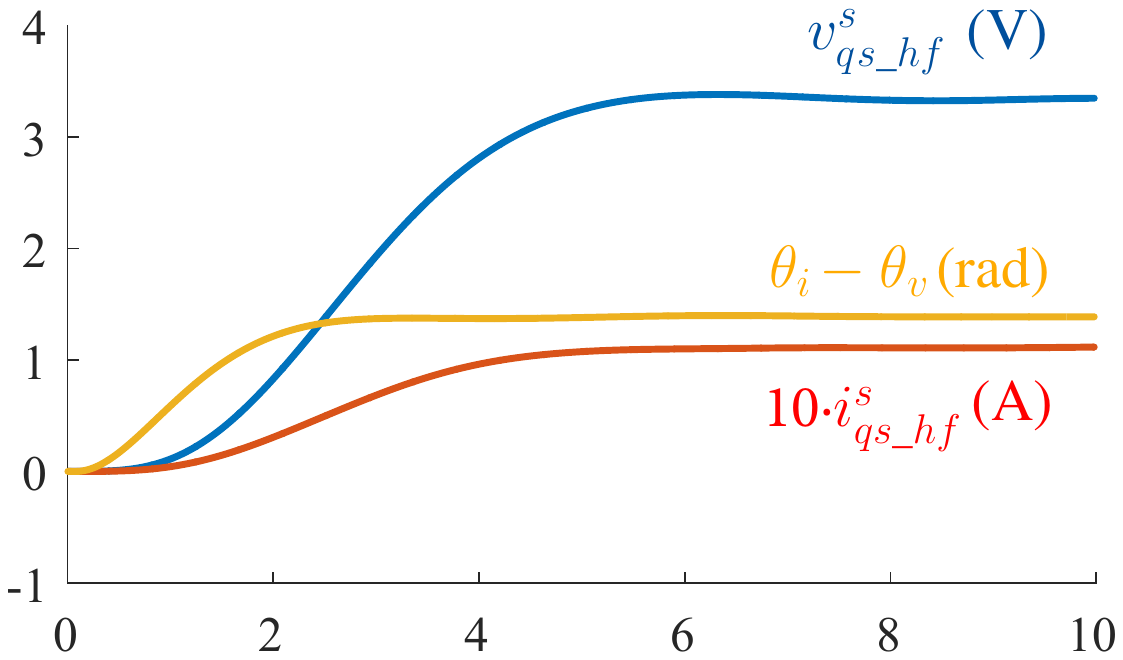}}
\subfloat{\includegraphics[width=2.2in]{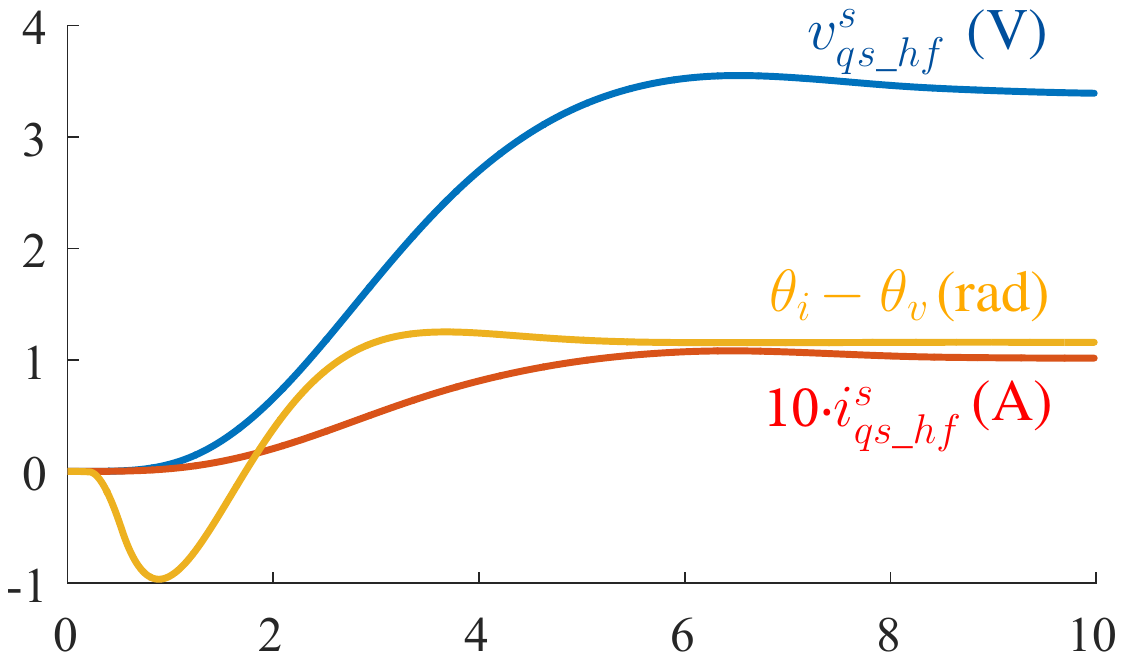}}
\\
\vspace{-0.125in}
\subfloat{\includegraphics[width=2.2in]{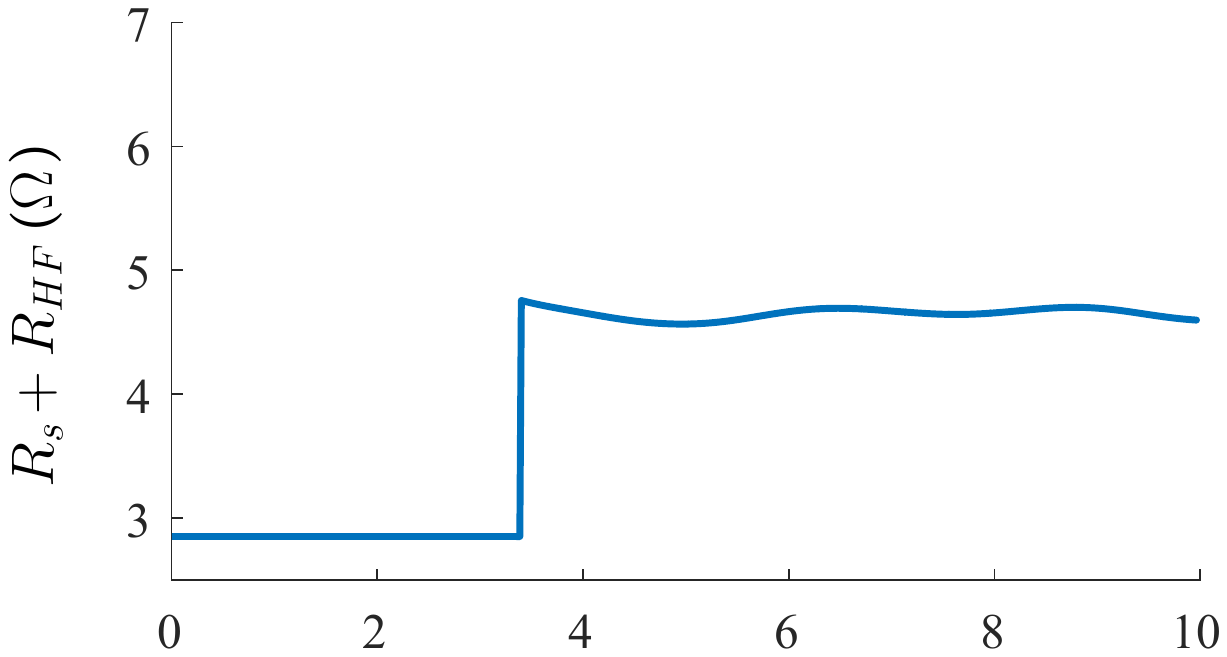}} 
\subfloat{\includegraphics[width=2.2in]{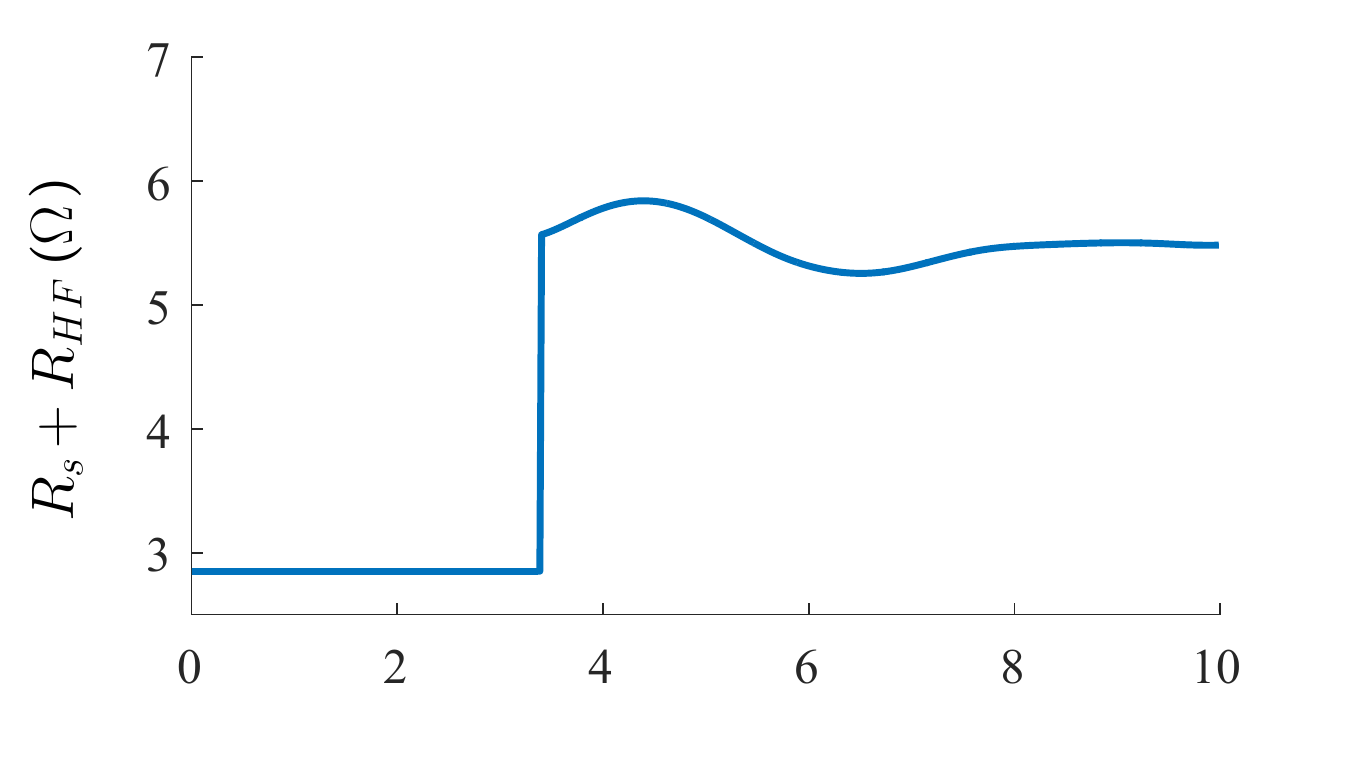}}
\subfloat{\includegraphics[width=2.2in]{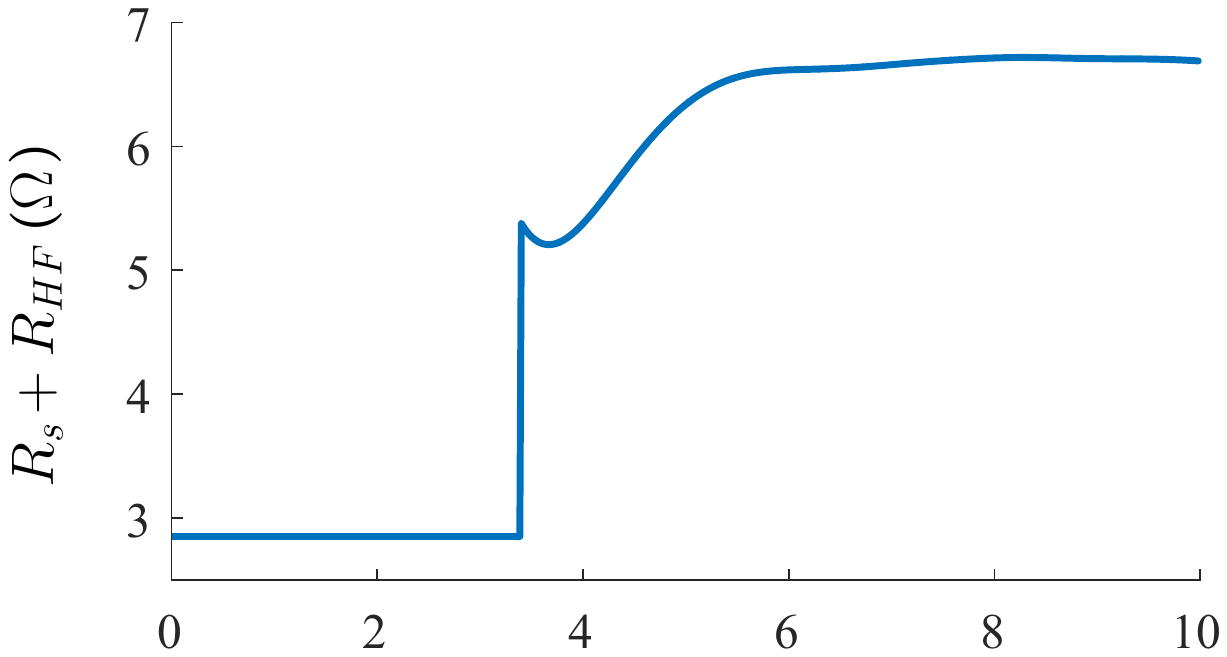}}\
\\
\subfloat[]{\includegraphics[width=2.2in]{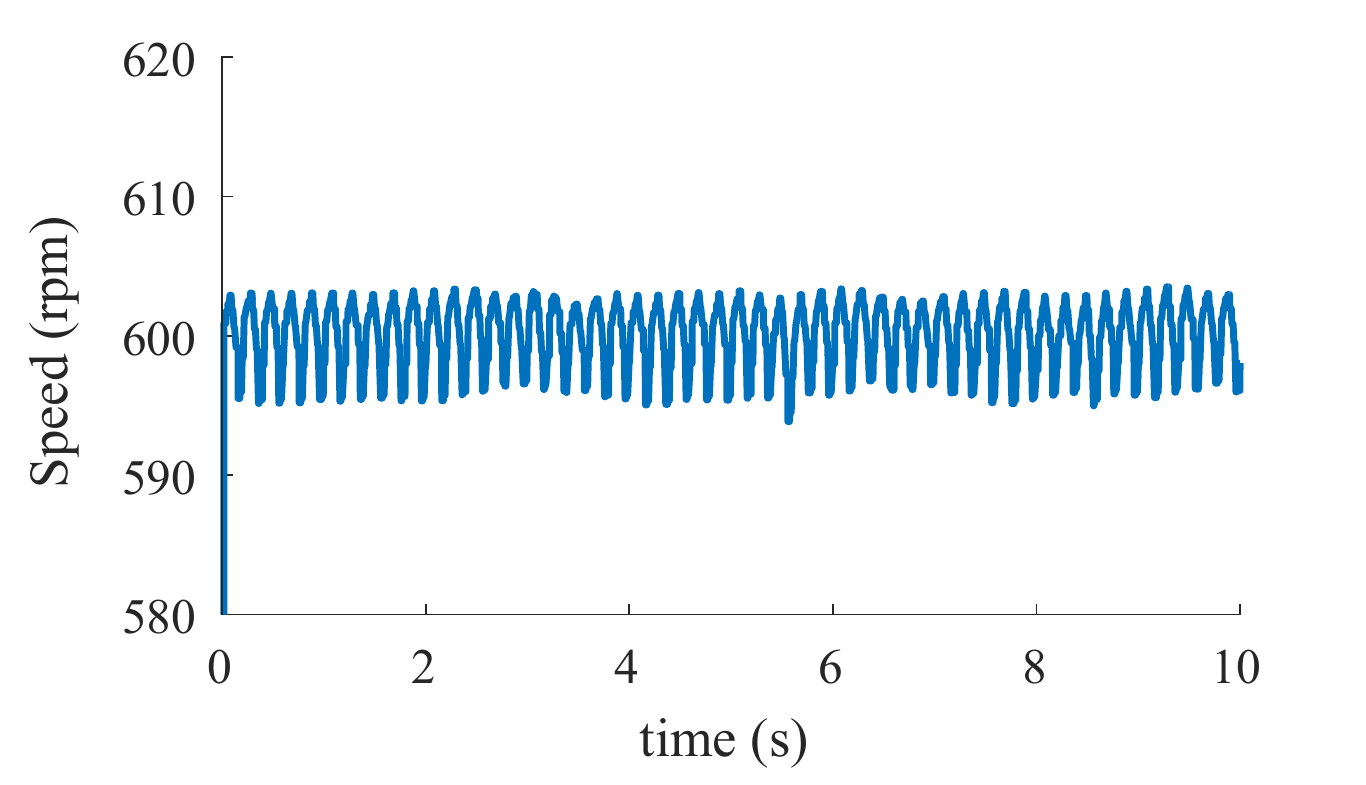}} 
\subfloat[]{\includegraphics[width=2.2in]{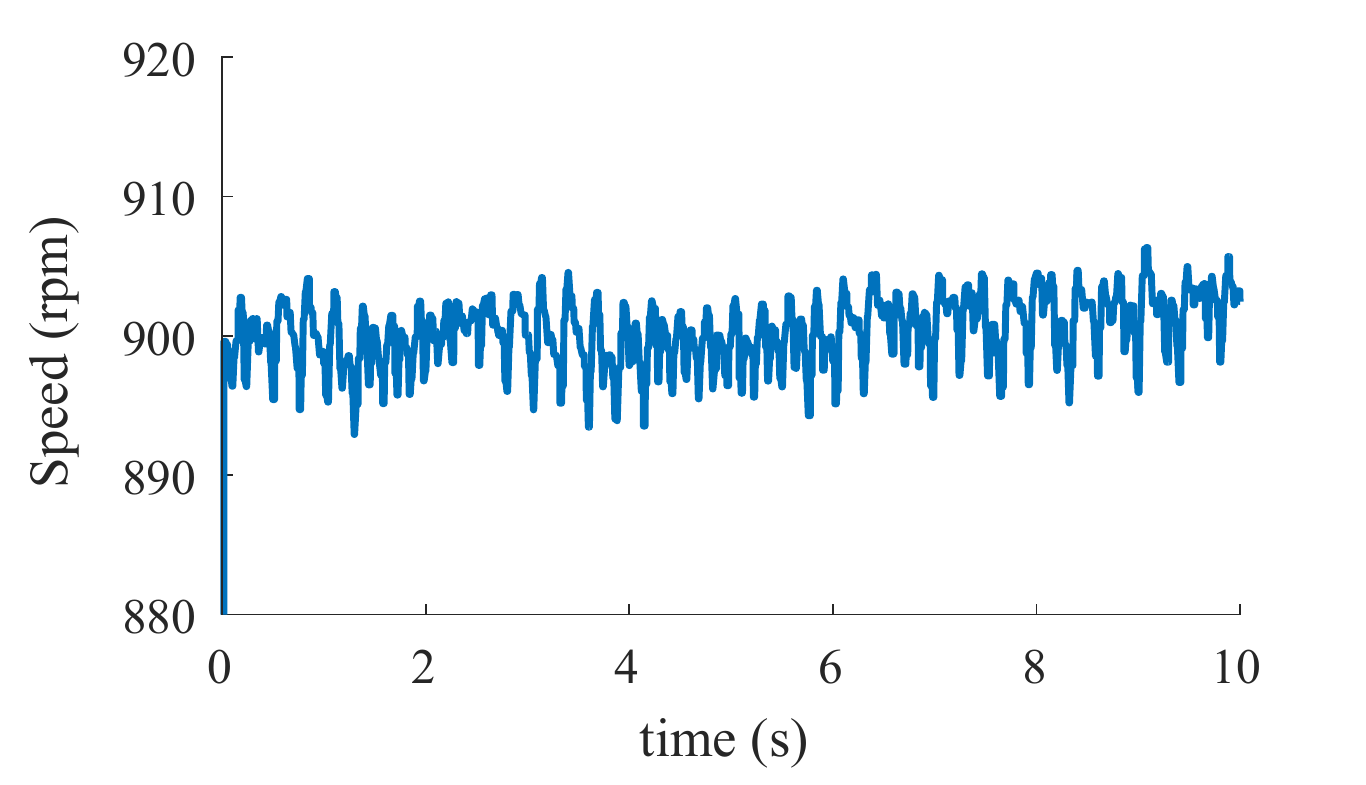}}
\subfloat[]{\includegraphics[width=2.2in]{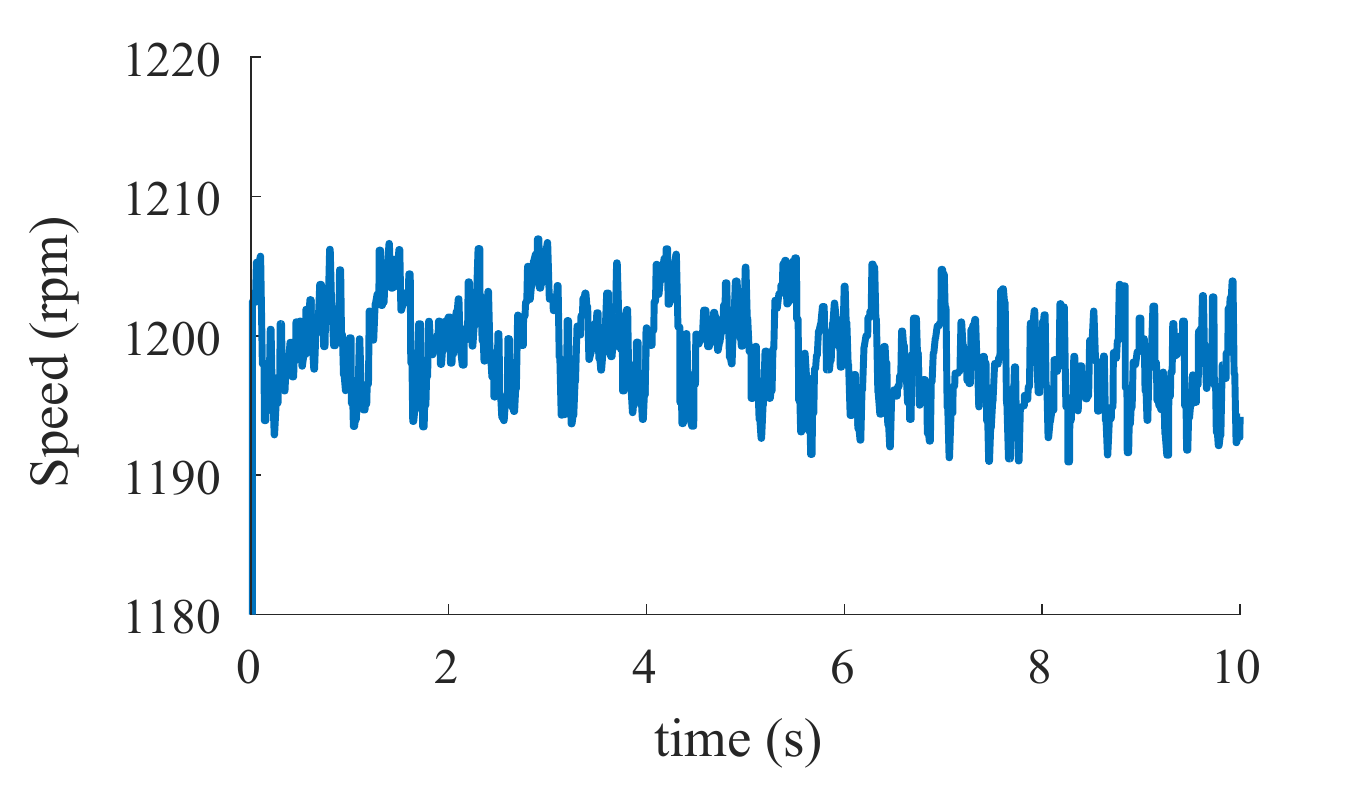}}
\caption{High-frequency voltage, current, phasor angle, resistance, and speed signals with high-frequency rotating flux injection at (a) 600 rpm; (b) 900 rpm and (c) 1,200 rpm.} 
\label{fig:PM_flux_VIR} 
\end{figure*}

The high-frequency voltage, current, phasor angle, resistance, and speed signals are presented in  with high-frequency rotating flux injection Fig. \ref{fig:PM_flux_VIR} at (a) 600 rpm; (b) 900 rpm and (c) 1,200 rpm. Because of their small magnitudes, the high-frequency current signals are magnified by 10 times to comfortably place the voltage, current and phasor offset values in one plot. It can be observed that for a typical 15 second injection period, the estimated resistance needs around 4 to 6 second to stabilize due to delays in the phasor extraction algorithm and the proposed real-time signal processing method demonstrated in Fig. \ref{fig_signal}. In addition, the estimated resistance values are larger at higher speeds, clearly indicating this reflected magnet resistance is largely induced by the eddy current effect on the permanent magnets, which would become more significant at higher frequencies.

\begin{figure*}
\centering
\subfloat{\includegraphics[width=2.2in]{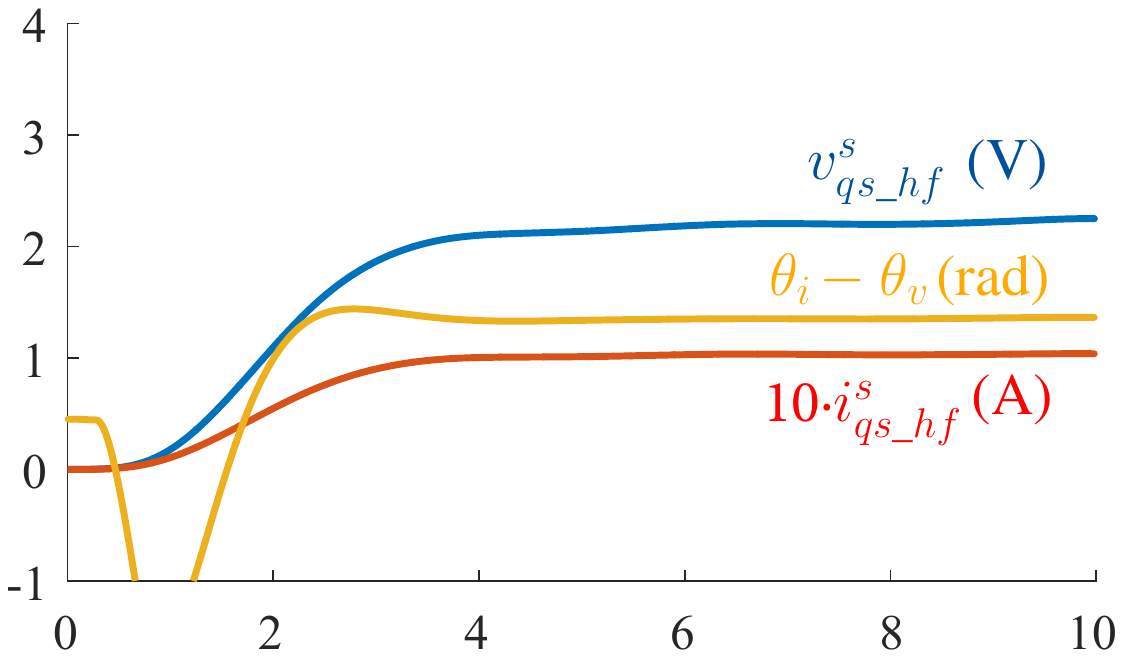}} 
\subfloat{\includegraphics[width=2.2in]{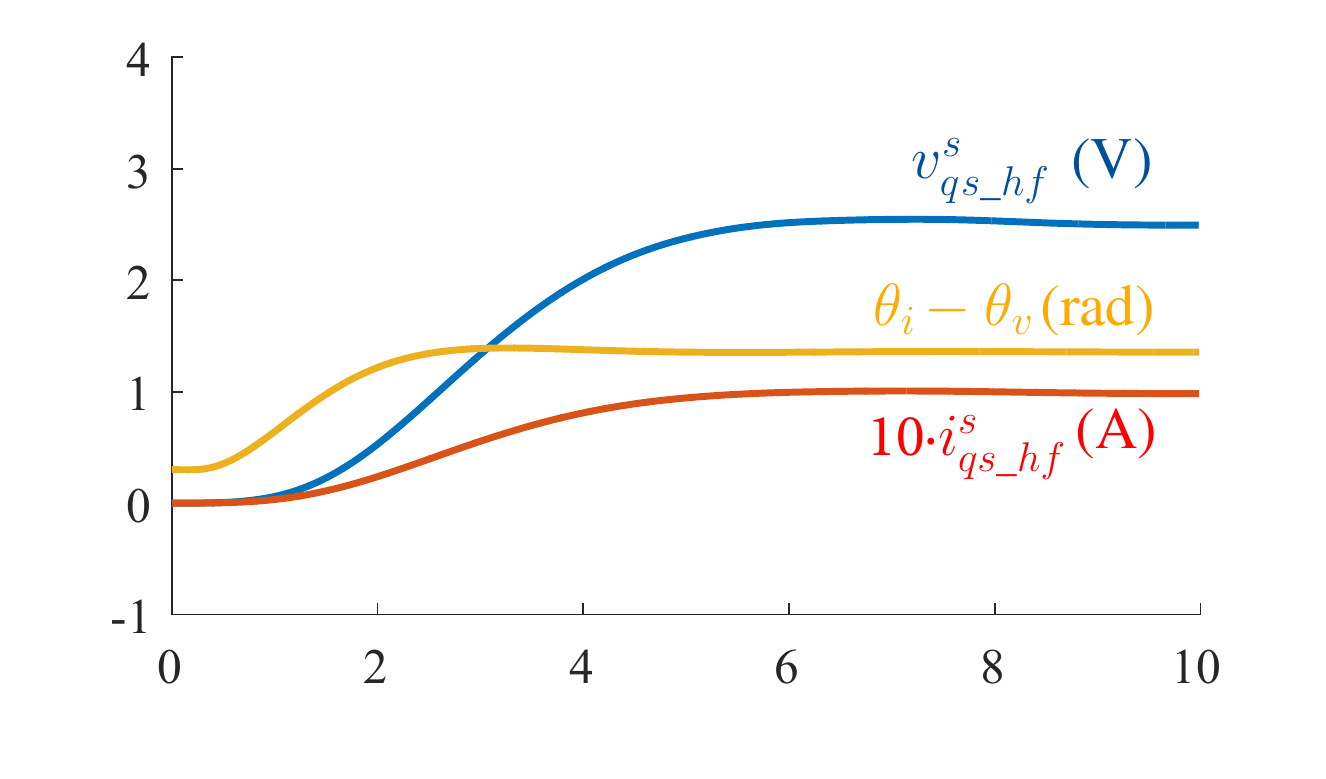}}
\subfloat{\includegraphics[width=2.2in]{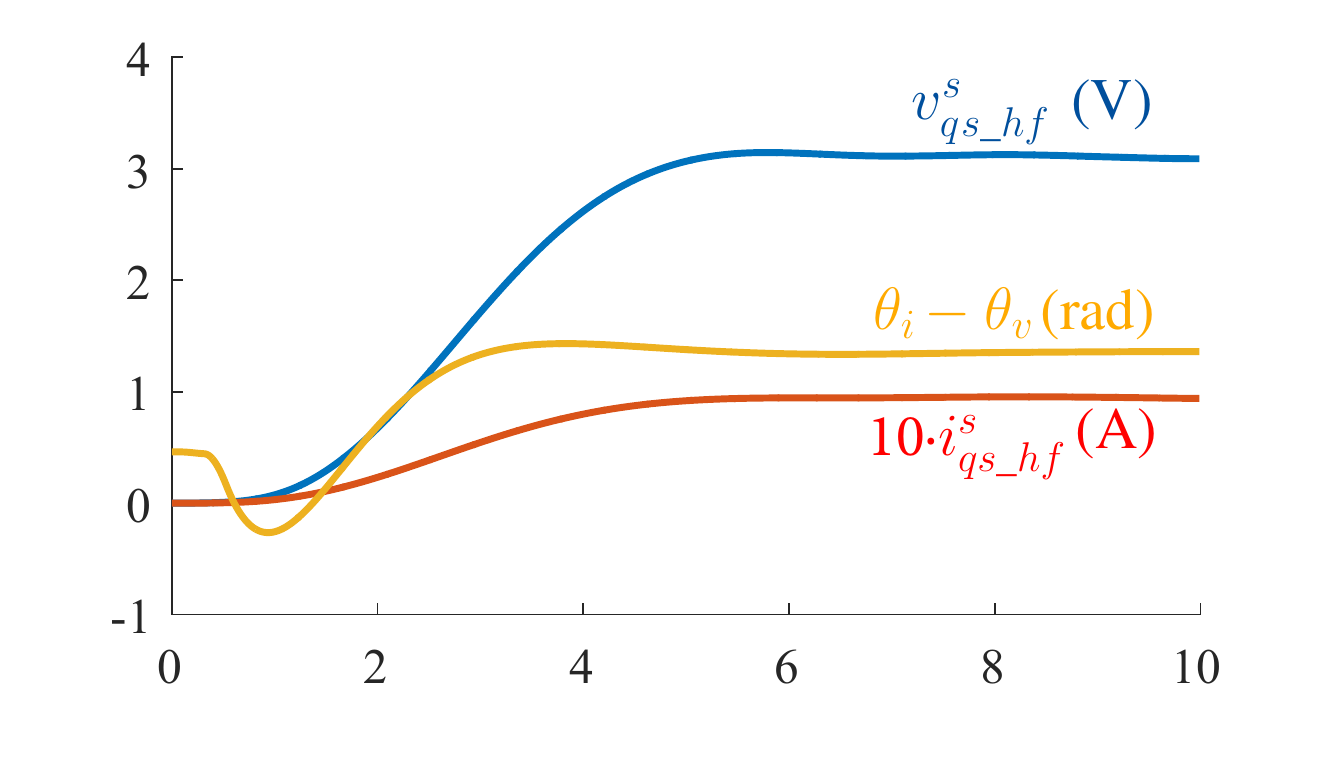}}
\\
\vspace{-0.125in}
\subfloat{\includegraphics[width=2.2in]{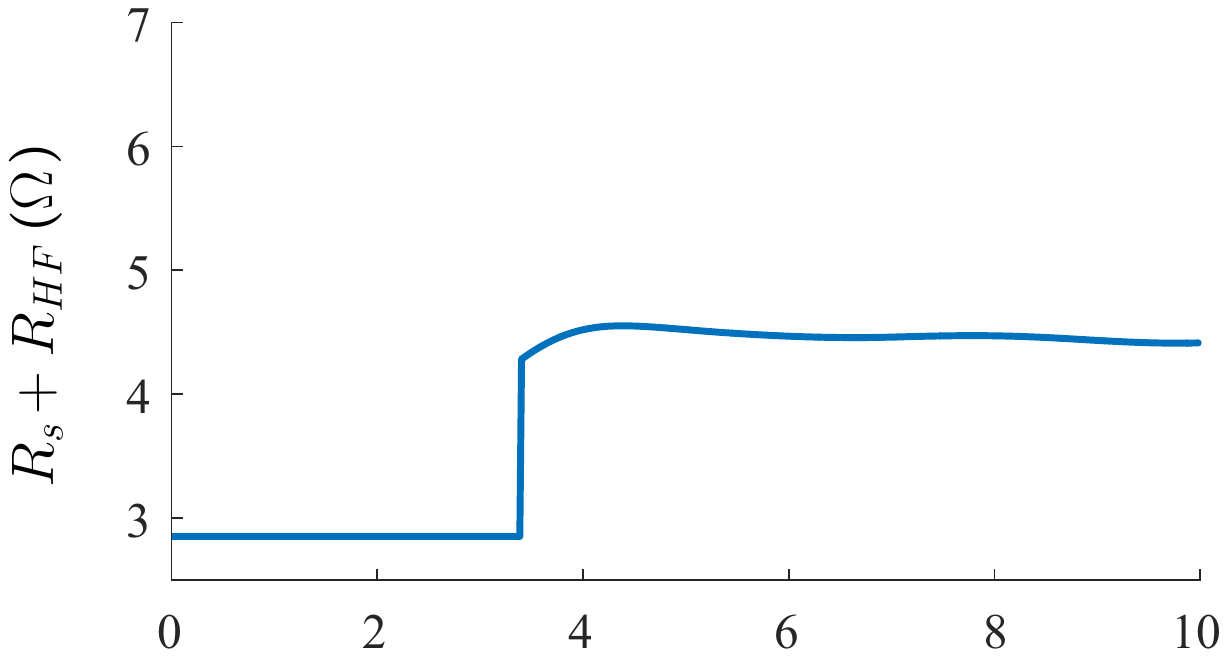}} 
\subfloat{\includegraphics[width=2.2in]{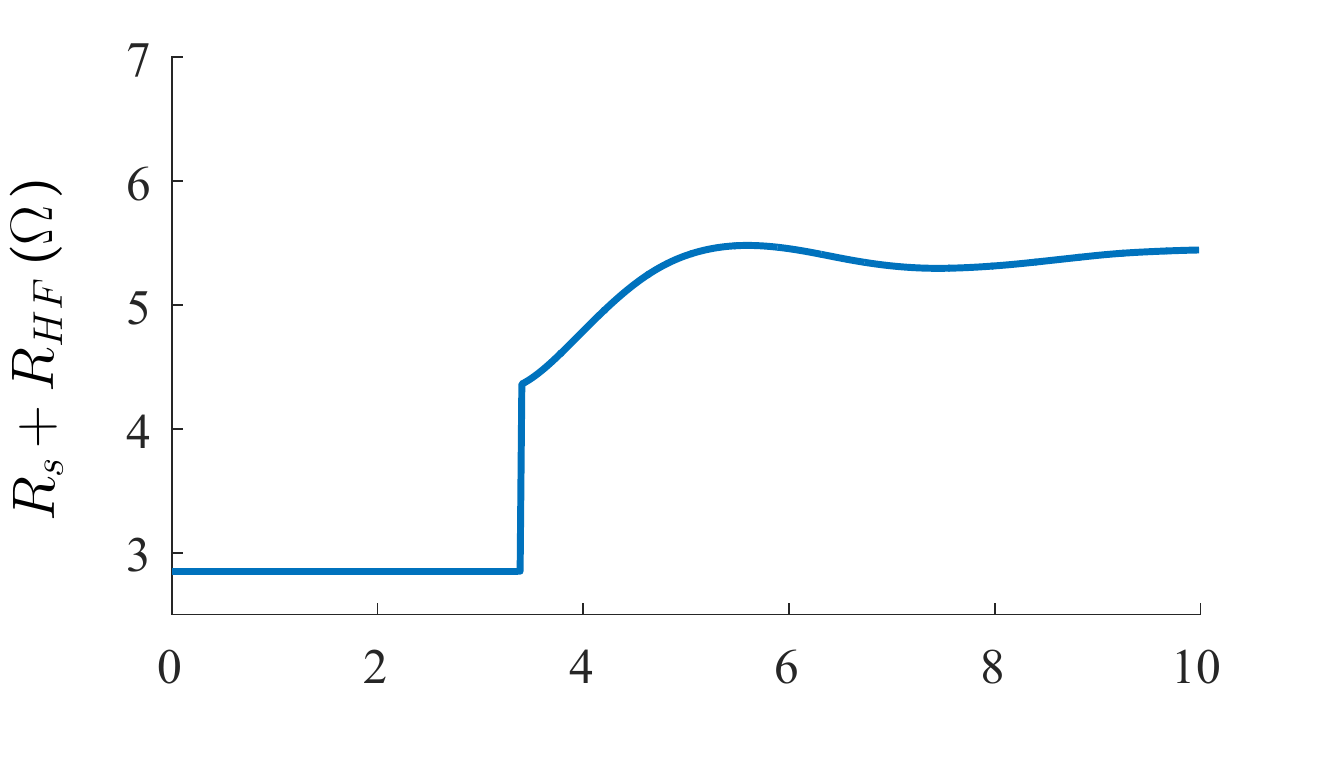}}
\subfloat{\includegraphics[width=2.2in]{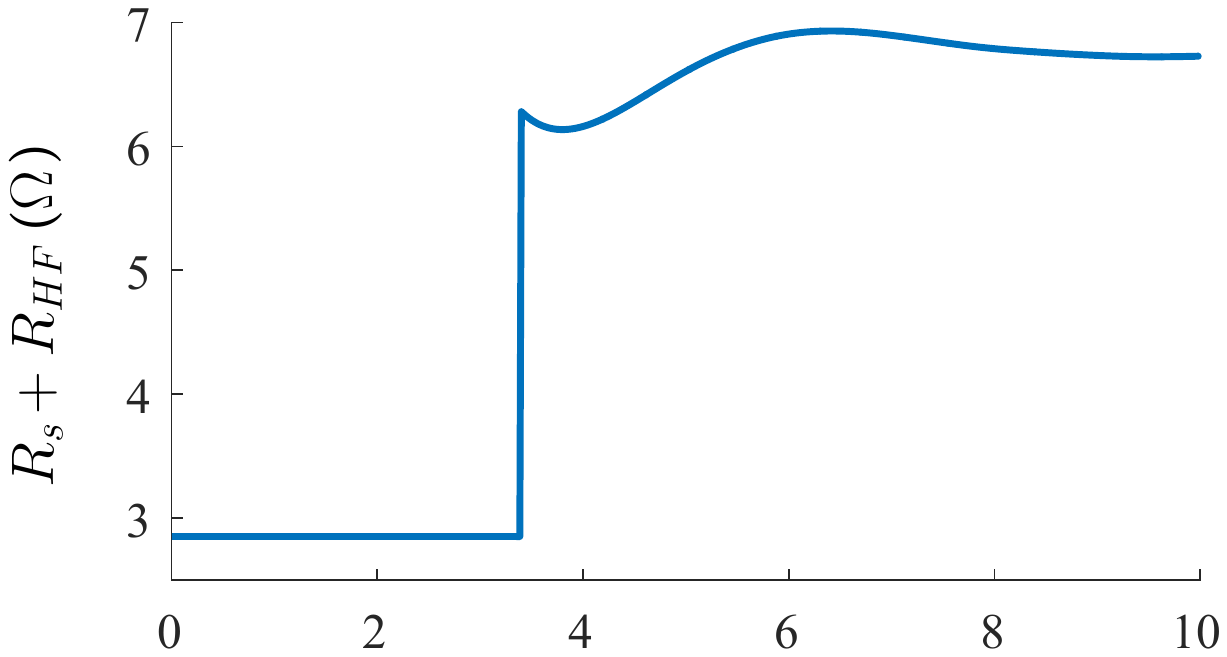}}
\\
\subfloat[]{\includegraphics[width=2.2in]{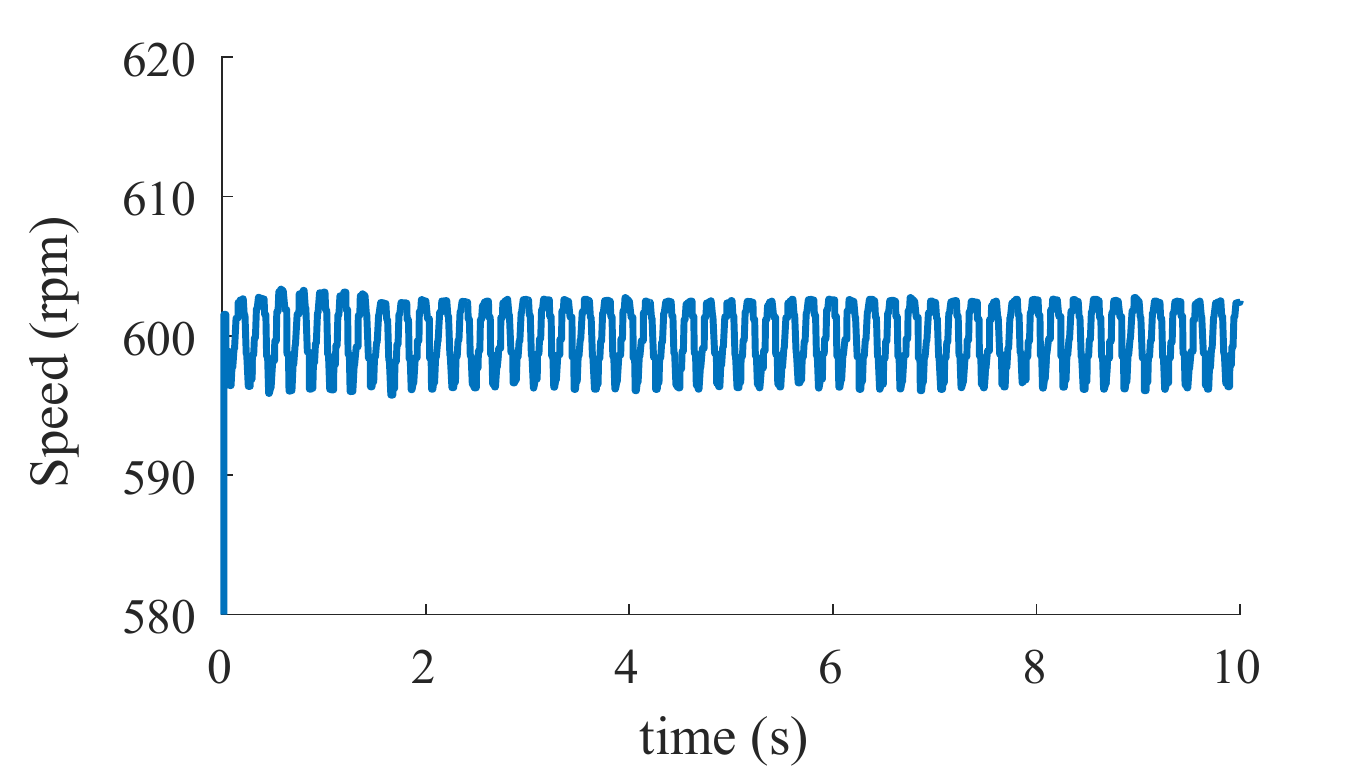}} 
\subfloat[]{\includegraphics[width=2.2in]{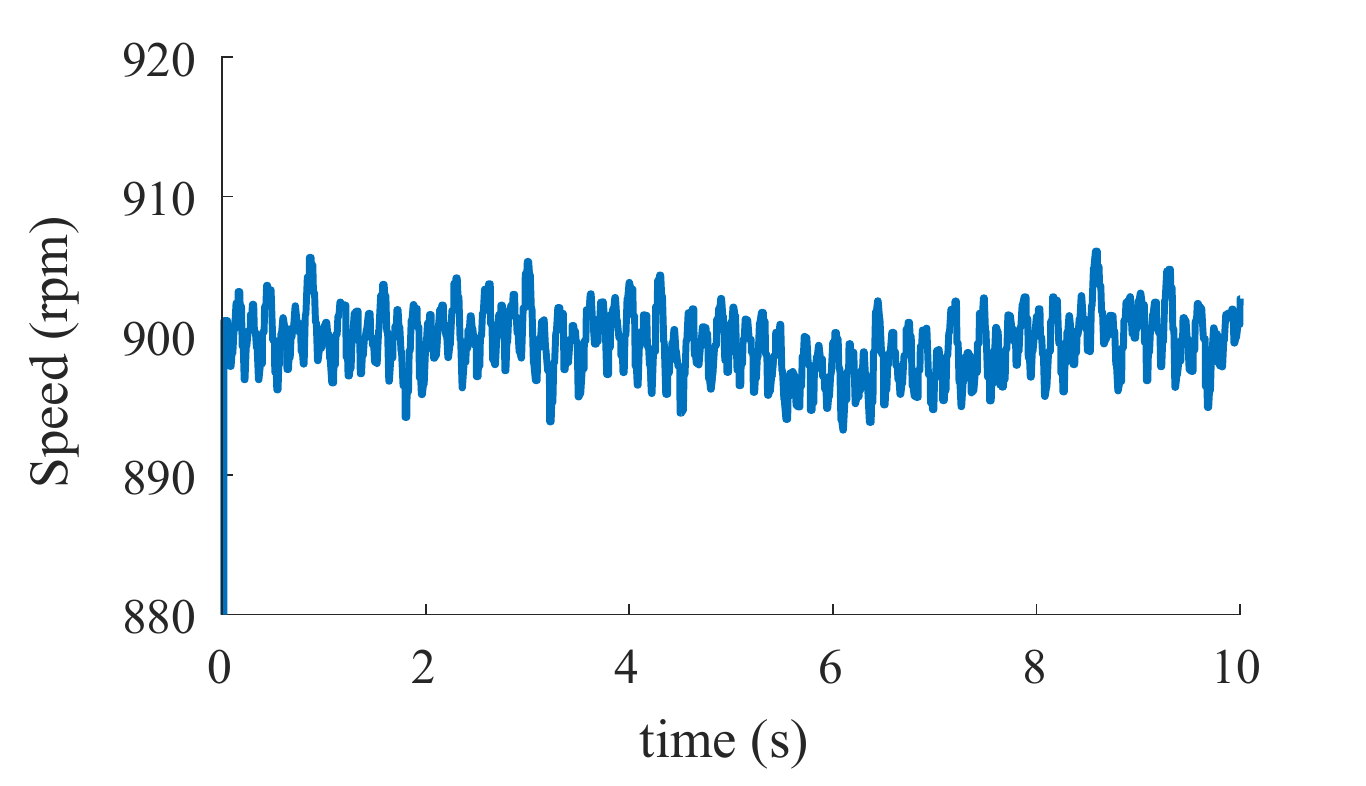}}
\subfloat[]{\includegraphics[width=2.2in]{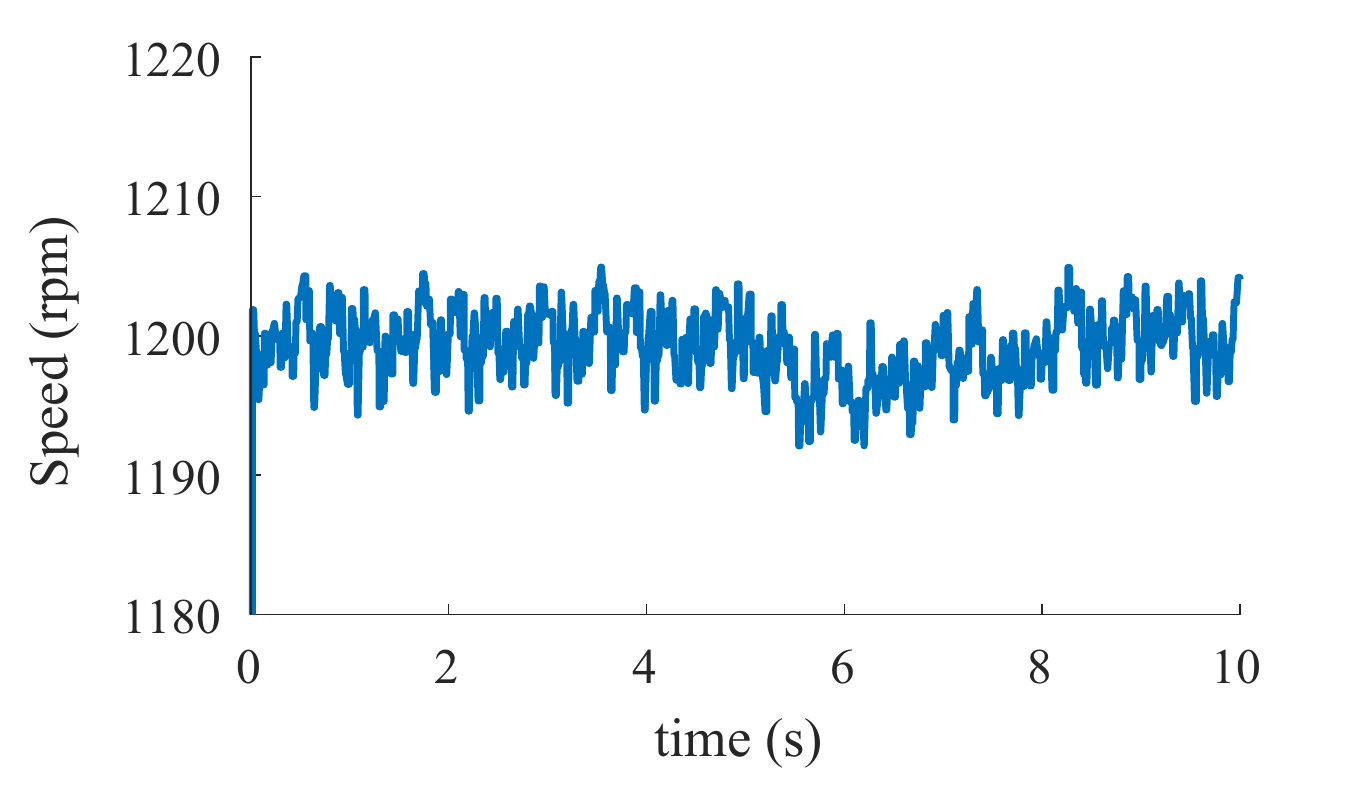}}
\caption{High-frequency voltage, current, phasor angle, resistance, and speed signals with high-frequency torque injection at (a) 600 rpm; (b) 900 rpm and (c) 1,200 rpm.} 
\label{fig:PM_torque_VIR} 
\end{figure*}

Very similar results and resistance values are estimated with the proposed high-frequency torque injection scheme, and are shown in Fig. \ref{fig:PM_torque_VIR}.

\begin{figure}[!t]
\centering
\subfloat[]{\includegraphics[width=3in]{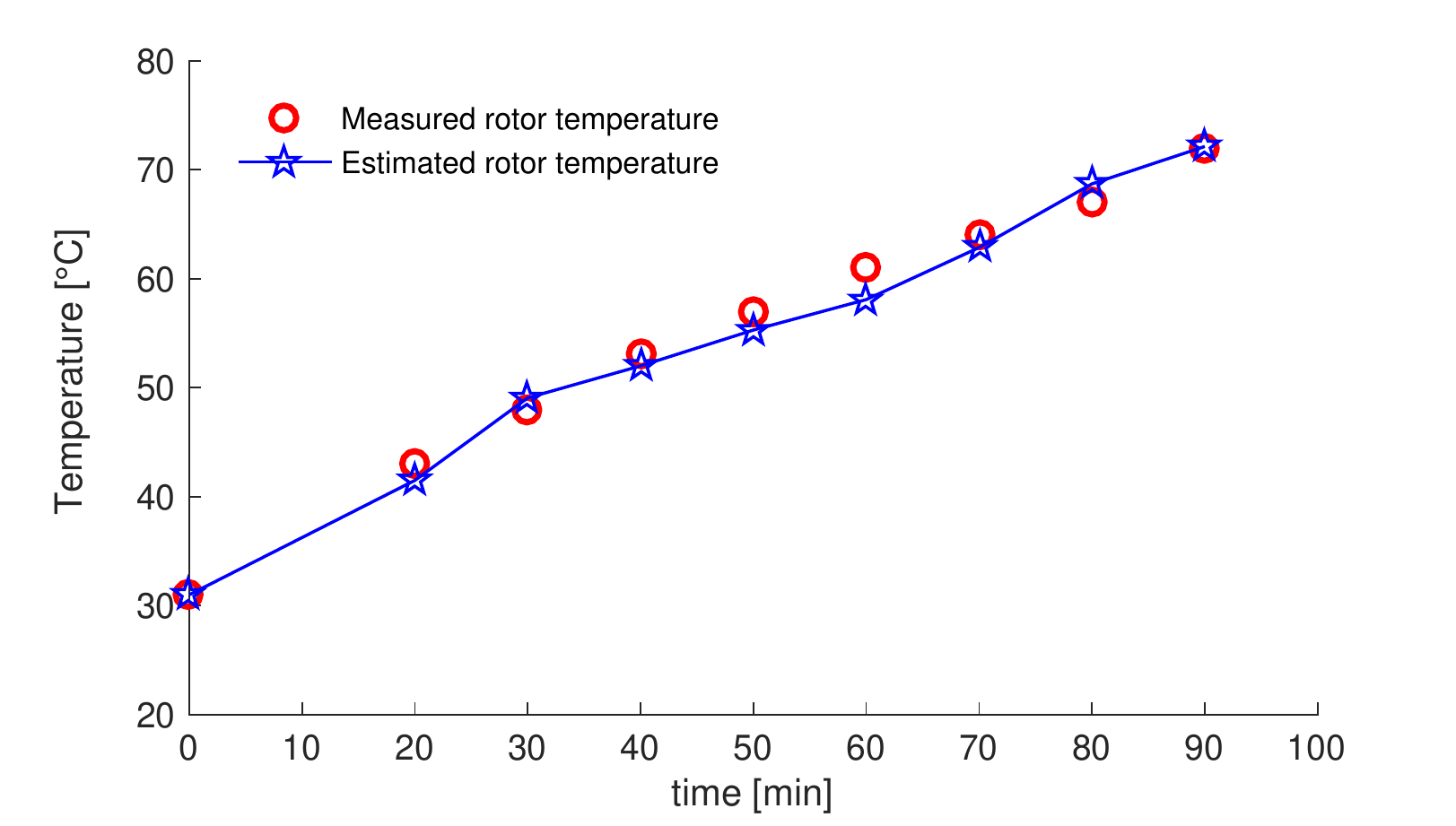}} 
\caption{Rotor temperature estimation performance with high-frequency rotating flux injection.} 
\label{fig:PM_flux_temp} 
\end{figure} 
\begin{figure}[!t]
\centering
\subfloat[]{\includegraphics[width=3in]{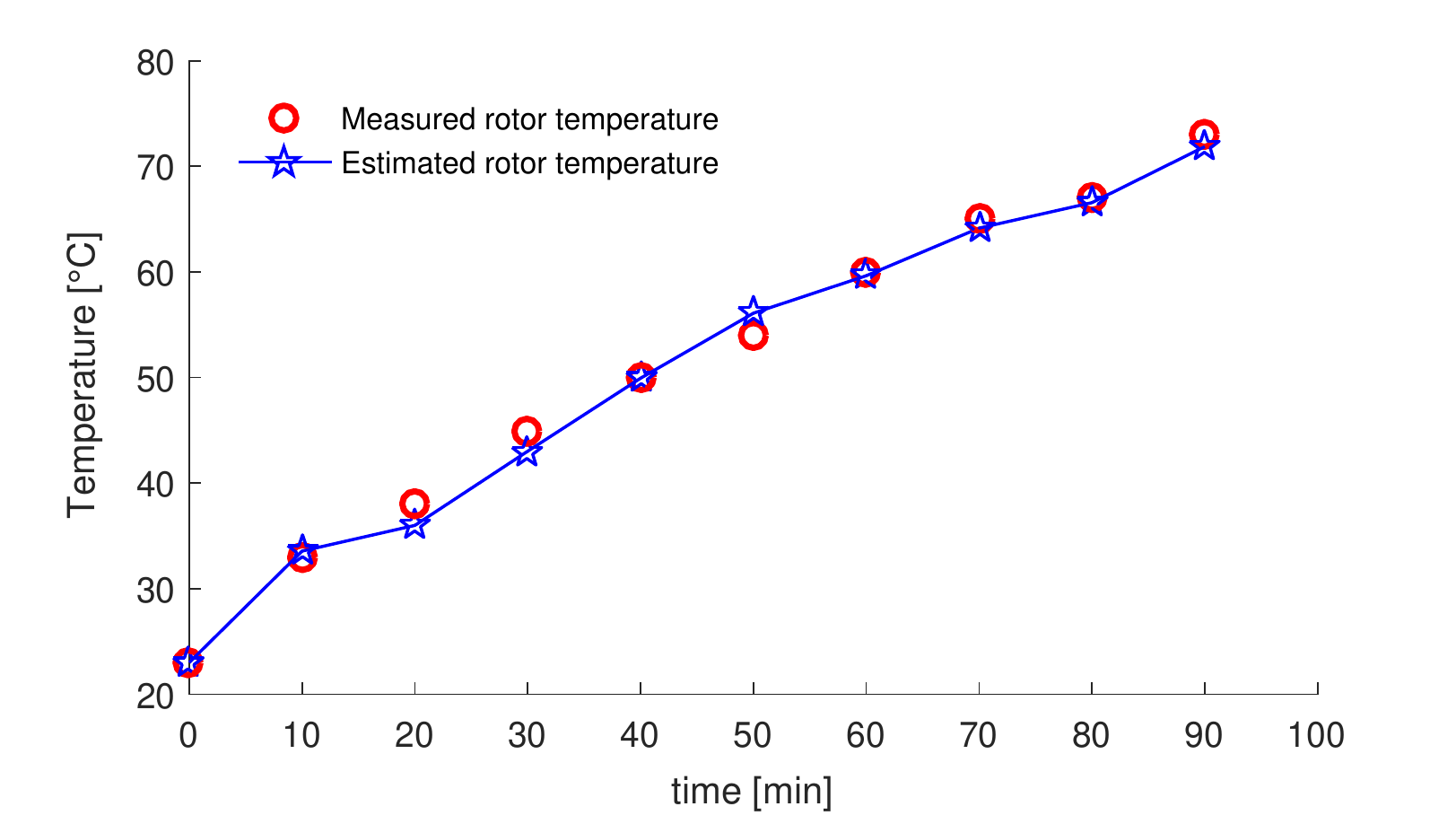}} 
\caption{Rotor temperature estimation performance with high-frequency torque injection.} 
\label{fig:PM_torque_temp} 
\end{figure} 

\subsection{Temperature Estimation at Constant Load }
To validate the proposed method, the IPM machine is running continuously at 150\% of its rated current at 600 rpm, and the torque reference is injected into the DTC algorithm for 15 seconds every 10 minutes. The reference stator resistance $R_{s0}$ is first measured when the machine is at the room temperature $T_0$. Then the PM temperature is tracked over time using Eqn. (4). The experimental results with both the measured PM temperature and the estimated PM temperature of the flux injection and torque injection are shown in Fig. \ref{fig:PM_flux_temp} and Fig. \ref{fig:PM_torque_temp}, respectively. As can be observed, the estimated magnet temperature tracks the temperature measurement from the infrared sensors very closely. The maximum temperature deviation is less than 3 \textcelsius~for both plots.

\section{Conclusion}
This paper proposes a nonintrusive thermal monitoring scheme for the permanent magnets inside the direct-torque-controlled IPM machines using high-frequency rotating flux or torque injection. The proposed method requires no additional sensors or hardware except for those already available in the IPMSM drives. Compared to the temperature measurement taken by the infrared sensors, the maximum error of the proposed method is less than $3^\circ C$ at a constant load condition.


\bibliographystyle{IEEEtran}
\balance
\bibliography{IEEEabrv_Shen.bib,ref.bib} 
%
%
%

\end{document}